\documentclass[prl,twocolumn,aps,preprintnumbers,nofootinbib]{revtex4-2}
\usepackage{graphicx}
\usepackage{color}
\usepackage{bm}
\usepackage{amsmath}
\usepackage{amssymb}
\usepackage{xspace}
\usepackage[normalem]{ulem}
\usepackage{units}
\usepackage{dcolumn}
\usepackage{paralist}
\usepackage{natmove}
\usepackage{natbib}
\usepackage{hyperref}

\begin{document}

\title{Investigation of
Phonon Lifetimes and Magnon-Phonon Coupling in YIG/GGG Hybrid Magnonic Systems in the Diffraction Limited Regime}

\author{Manoj~Settipalli}
\affiliation{Ann and H.J. Smead Aerospace Engineering Sciences, University of Colorado Boulder, Boulder, Colorado 80303, USA}

\author{Xufeng Zhang}
\affiliation{Electrical and Computer Engineering, Northeastern University, Boston, MA 02115}

\author{Sanghamitra Neogi}
\email{sanghamitra.neogi@colorado.edu}
\affiliation{Ann and H.J. Smead Aerospace Engineering Sciences, University of Colorado Boulder, Boulder, CO 80303, USA}

%\date{\today}

\begin{abstract}
Quantum memories facilitate the storage and retrieval of quantum information for on-chip and long-distance quantum communications. Thus, they play a critical role in quantum information processing and have diverse applications ranging from aerospace to medical imaging fields. Bulk acoustic wave (BAW) phonons are one of the most attractive candidates for quantum memories because of their long lifetime and high operating frequency. In this work, we establish a modeling approach that can be broadly used to design hybrid magnonic high-overtone bulk acoustic wave resonator (HBAR) structures for high-density, long-lasting quantum memories and efficient quantum transduction devices. We illustrate the approach by investigating a hybrid magnonic system, where BAW phonons are excited in a gadolinium iron garnet (GGG) thick film via coupling with magnons in a patterned yttrium iron garnet (YIG) thin film. We present theoretical and numerical analyses of the diffraction-limited BAW phonon lifetimes, modeshapes, and their coupling strengths to magnons in planar and confocal YIG/GGG HBAR structures. We utilize Fourier beam propagation and Hankel transform eigenvalue problem methods and discuss the effectiveness of the two methods to predict the HBAR phonons. We discuss strategies to improve the phonon lifetimes, since increased lifetimes have direct implications on the storage times of quantum states for quantum memory applications. We find that ultra-high, diffraction-limited, cooperativities and phonon lifetimes on the order of ${\sim}10^5$ and ${\sim}10$ milliseconds, respectively, could be achieved using a CHBAR structure with $10\mu$m lateral YIG dimension. Additionally, the confocal HBAR structure will offer more than 100-fold improvement of integration density. A high integration density of on-chip memory or transduction centers is naturally desired for high-density memory or transduction devices. Our results will have direct applicability for devices operating in the cryogenic or milliKelvin regimes. For example, our approach and analyses can be applied to design HBAR devices that could effectively couple with superconducting qubit systems. 

\end{abstract}

\maketitle

\section{Introduction}

Hybrid quantum systems that integrate diverse platforms, constitute a rapidly emerging research field due to their ability to synergistically address the limitations of individual platforms. Among various systems explored, hybrid magnonic systems received significant attention in recent years due to the unique advantages they offer~\cite{lachance2019hybrid,li2020hybrid}. In these systems, quantized spin waves (magnons) coherently couple with other information carriers, such as photons and phonons. The coupling presents avenues for both fundamental investigations and practical implementations  ~\cite{Huebl2013,Zhang2014,Tabuchi2014,Bai2015,Goryachev2014,Zhang2016,Osada2016,Haigh2016,Zhang2016a}. These systems often leverage magnetic materials with high spin density, such as yittrium iron garnet (YIG). Such materials enable strong coupling of magnons with microwave photons trapped in a cavity, with coupling strengths surpassing their respective dissipation rates~\cite{Huebl2013,Zhang2014,Tabuchi2014,Bai2015,Goryachev2014}. Several devices have been investigated that benefit from the strong magnon-photon coupling, with applications ranging from microwave-to-optical transduction~\cite{Hisatomi2016,Zhu2020a} to quantum magnonics~\cite{Tabuchi2015,Lachance-Quirion2020,Wolski2020}, dark matter detection~\cite{Flower2019,Trickle2020} and others~\cite{Wang2019, Zhang2020, Zhu2020,Zhang2017,Harder2017,Zhang2019}.  

Besides magnon-photon coupling, magnon-phonon coupling enables another important class of hybrid magnonic devices which has attracted attention for coherent and quantum information processing applications~\cite{Zhang2016a,bozhko2017bottleneck,lisenkov2019magnetoelastic,verba2019wide,shah2020giant}. Phonons are particularly attractive due to their long lifetime compared to other information carriers~\cite{Li2008, Weis2010, Teufel2011, Chan2011}. Magnon-phonon coupling has shown success for classical information processing applications~\cite{kittel1958,schlomann1960,eshbach1963,schlomann1964,damon1965dispersion,auld1967adiabatic,comstock1967magnetoelastic,rezende1969magnetoelastic,yukawa1974fmr,camley1979magnetoelastic,komoriya1979magnetoelastic,zaitsev1998magnetoacoustic,filimonova2004nonlinear}. The coupling of magnons with mechanical phonons in YIG spheres has been utilized for promising quantum information processing (QIP) applications, such as magnon-phonon entanglement~\cite{li2019entangling}, nonreciprocal phonon propagation~\cite{zhang2020unidirectional,xu2020nonreciprocal}, and magnon squeezing~\cite{Li2023May}. Separately, recent studies used high-overtone bulk acoustic wave resonator (HBAR) phonons, that operate in the GHz regime, for coupling with superconducting qubits and demonstrated their remarkable potential for quantum acoustodynamics~\cite{Chu2018Nov,renninger2018bulk,Gokhale2020May}. The HBAR phonons also exist in hybrid magnonic devices and can enable resonant magnon-phonon coupling via the magnetoelastic effect~\cite{Vanderveken2020}. These phonons are commonly supported by structures composed of YIG thin-films~\cite{ordonez2007observation,polzikova2016acoustic,Chowdhury2017,an2020coherent} grown on a thick gadolinium gallium garnet (GGG) substrate, which can host a series of HBAR resonances. A recent study demonstrated a triply resonant photon-magnon-phonon system using a YIG/GGG hybrid magnonic HBAR device~\cite{xu2021coherent}.
% reported coherent pulse echoes in , demonstrating . 
Such resonant coupling can enable long-lived multimode phononic quantum memories and other quantum information processing and transduction applications. In this article, we investigate a class of planar and confocal HBAR YIG/GGG hybrid magnonic structures for quantum transduction applications.

The performance figure of merit for magnon-phonon transduction can be characterized using cooperativity, $C=4g_{mb}^2/\kappa_m\kappa_b$, where $g_{mb}$, $\kappa_m$, and $\kappa_b$ are magnon-phonon coupling strength, magnon dissipation rate, and phonon dissipation rate, respectively. The lifetimes of magnons and phonons are given by $\tau_m=1/\kappa_m$ and $\tau_b=1/\kappa_b$, respectively. In YIG/GGG hybrid magnonic HBAR devices, the phonon lifetimes usually surpass the magnon lifetime thanks to the excellent mechanical properties of the garnets. Past studies reported the magnon and phonon lifetime in YIG/GGG HBAR systems to be $\tau_m=0.07\mu$s and $\tau_b~\sim0.25\mu$s, respectively~\cite{an2020coherent,xu2021coherent}. The system lifetime is primarily determined by the phonon lifetimes. It is desirable to further improve the phonon lifetimes and consequentially, the system lifetimes, for quantum memories and other information processing applications. However, a complete understanding about the loss mechanisms that limit the phonon lifetime in garnet devices is not yet available. The phonon lifetime could be limited by material and diffraction losses depending on the device geometry or the operating conditions. At room temperature, the phonon lifetime is primarily limited by acoustic attenuation due to phonon-phonon interactions~\cite{dutoit1972ultrasonic,dutoit1974microwave}. The acoustic attenuation effects, $\alpha \propto 1/\tau_b$, follows a $T^4$ temperature dependence. Due to the $T^4$ behavior, these effects are less significant at low temperatures, such as millikelvin regime where many qubit systems operate. Ideally, $\tau_b$ can increase by multiple orders of magnitude at cryogenic temperatures. However, the diffraction losses play an important role in determining the phonon lifetime at low-temperature operating conditions, where the material losses are suppressed. In this study, we establish a computational approach for analyzing the diffraction losses in HBAR structures and use it to investigate the effects of diffraction losses on the HBAR phonons of hybrid YIG/GGG magnonic structures. We focus on the diffraction effects because they can be analyzed computationally, while experimental characterization is needed to investigate the acoustic attenuation effects. Note that the knowledge of material losses in YIG/GGG material systems is limited. There is a strong need for experimental characterization of acoustic attenuation in different magnetic materials across temperature and operating frequencies, to unlock their full potential.

%Following the previous arguments, if diffraction losses are weaker than the material losses, one could achieve multiple orders of magnitude improvement in $\tau_b$ as one goes to cryogenic temperatures due to the $T^4$ behavior. Therefore, it is important to characterize the diffraction losses of planar HBAR systems to understand if low-T operations are a viable solution to generate long-lifetime phonons. 

% Although one could enhance the material-limited lifetime by operating in the mK temperature regime, 

The integration density of the hybrid magnonic HBAR structures is an important design aspect for developing high-density phononic quantum memories. The integration density is measured by the number of memory or transduction components that could be incorporated into a single chip, in addition to other on-chip circuitry. The YIG film represents the transduction component of the YIG/GGG HBAR structures of interest. Thus, the smaller the lateral dimension of the YIG film, the greater the number of transduction components in a single chip of given dimension and the integration density. We define the integration density as $D_i=\frac{A^0_d}{A_d}$. Here, $A_d$ is the lateral area of the YIG film for planar HBAR structures, and the cross-sectional area of the confocal dome surface for CHBAR structures in our study, respectively. $A^0_d$ represents a reference value for the device. We consider the smallest known lateral area reported for YIG/GGG devices in literature to be the reference, $A^0_d=0.8\times0.9$ \text{mm}$^2 = 0.72$ mm$^2$~\cite{xu2021coherent}. The integration density of the device can be increased by reducing the lateral dimension of the YIG films. However, the reduced aperture will lead to high-diffraction and affect phonon lifetime. In addition, a smaller aperture will reduce the overlap between the magnon and the phonon modes since they only overlap in the YIG film. A reduced magnon-phonon overlap could lead to a weaker magnon-phonon coupling strength. Consequently, the cooperativity will decrease as a result of the reduced lifetime and coupling strength. It is therefore imperative to develop a strategy to increase the integration density without inducing excessive diffraction losses. Note that we only focus on the lateral integration density while keeping the thickness fixed at $527.2\mu$m for all structures investigated in this work. The same thickness allows us to keep the free spectral range of phonons fixed for all our analysis.
  
%Another important aspect in designing such a device/structure is its integration density, which is characterized by the lateral area of the YIG film. The smaller this area, the greater the number of hybrid magnonic memory or transduction stations that could be incorporated into a single chip, in addition to other on-chip circuitry. Most YIG/GGG systems only mention their device thickness dimensions, hence for this study, the reference for YIG film's lateral dimensions is obtained from Ref.\cite{xu2021coherent}, which is $0.8\times0.9$ \text{mm}$^2$ making this is a millimeter-scale device. In order to achieve improved integration density, it is imperative to understand the performance of the structure when the lateral area is reduced. It is expected that a decrease in the lateral dimensions of the YIG film will lead to high-diffraction due to reduced apertures. 

In this study, we investigate a set of planar and confocal YIG/GGG HBAR structures. We investigate the phonon lifetime and the magnon-phonon coupling strength in these structures, and identify strategies to improve their performance and integration density. We assume a magnon lifetime of $0.07\mu$s~\cite{xu2021coherent} in all structures considered in this study. The phonon lifetime $\tau_b$ or the quality factor ($Q=\omega\tau_b$) is of particular interest due to its direct relevance to the quantum memory storage time. 
% Generally, studies investigating mechanical resonators report quality factors ($Q=\omega\tau_b$) to quantify the quality of the phonon modes. 
%Due to their direct proportionality, $Q$-factor and $\tau_b$ refer to the same property of the phonon mode at a given frequency $\omega$. 
Henceforth, we drop the subscript $b$ from $\tau_b$, unless otherwise needed to distinguish it from magnon and photon lifetimes. Without losing generality, we only consider the fundamental magnon mode, i.e., the Kittel mode, for simplicity. The Kittel mode has a well-defined analytical function for circular discs. However, such analytical description is still lacking for the HBAR phonon modes. 
% We evaluate existing numerical methods and discuss the appropriate methods that can reliably model the HBAR phonons modes. 
Several methods have been implemented to model HBAR phonons and their lifetimes. For example, phonon lifetime in planar HBARs was calculated by decomposing the initial beam in a Bessel function basis and calculating the overlap of the reflected beam with the initial profile after each round trip~\cite{chu2017quantum}. However, the predicted lifetime was smaller than the experimental observations %, although on the same order of magnitude. 
because the effects of the lateral confinement induced by the transducer are not included. 
Finite element analysis (FEA) is another popular method to study HBAR phonons, using the COMSOL Multiphysics~\cite{multiphysics2013comsol} software. A recent study estimated the diffraction loss in an epitaxial planar HBAR structure by measuring the power received at the opposite end to that of the actuator~\cite{gokhale2021phonon}. 
% % They modeled the actuator as a time-varying displacement and varied actuator lateral areas. 
% However, they considered half a round trip 
% % and did not account for the fact that diffraction losses can vary after each round trip 
% and also ignored the effect of localization due to the transducer. In general, acoustic wave amplitudes decay non-exponentially with time due to the mode mismatch between the input acoustic beam profile and the phonon mode of interest. Hence, it is important to consider how transducer localization affects over multiple round trips to accurately determine the characteristic decay of the mode of interest, instead of only half a round trip. 
Another study used FEA to predict the modeshapes of a confocal HBAR for qubit coupling applications~\cite{chen2019engineering}. 
% They reported quality factors with perfectly matched layer (PML) boundaries and also demonstrated the resilience of CHBAR to tilts/non-parallelism compared to the planar HBAR. 
They considered phonon modes at low-overtones or long wavelengths, hence it was possible to perform three-dimensional (3D) FEA simulations for this system. However, as we will discuss later, it becomes intractable to simulate  bulk acoustic waves with overtone numbers $n{\sim}3000$ using FEA since sampling required for these small wavelength modes is on the order of $~{\sim} $ billion nodes. 
Moreover, all these studies discussed longitudinal HBAR phonons whereas in hybrid magnonic devices shear phonons exhibit much stronger coupling with magnon modes~\cite{an2020coherent,xu2021coherent,schlitz2022magnetization}. The FEA analysis becomes particularly expensive for shear phonons since they are not axi-symmetric. One cannot leverage the axi-symmetric two-dimensional (2D) FEA modelling available in COMSOL Multiphysics for these systems. 
% A past study simulated a 2D YIG/GGG HBAR system in COMSOL without leveraging axi-symmetry~\cite{xu2021coherent}. They ignored the different impacts of Gouy phase effects between 2D and 3D diffraction~\cite{feng2001physical}. It is necessary to consider such effects to accurately predict lifetimes and modeshapes. 
Due to these reasons, we explore other approaches to model shear phonon modes. We consider the Fourier beam propagation method (FBPM)~\cite{renninger2018bulk,chu2017quantum} which follows a Fox-Li-like~\cite{Fox1961} iterative approach to obtain the phonon modes and works with a plane-wave basis set. Although past studies used FBPM for longitudinal phonon modes, we illustrate that it can effectively model the shear phonon modes of interest. We provide a detailed description of this method and discuss an adaptive algorithm that allows to overcome some of the challenges of the standard FBPM method. Additionally, we consider % to the Fox-Li FBPM, we also consider another Fourier transform based method which leverages axi-symmetry of the structures. 
another method that uses Hankel transform (HT), which is a Fourier transfrom for axi-symmetric systems, and works with a Bessel function basis. Thus far, HT method has only been implemented for Fabry-Perot optical cavities~\cite{poplavskiy2018fundamental, vinet1993matrix}. We adapt this approach to simulate YIG/GGG HBAR structures by leveraging their axi-symmetry and isotropic material properties. 

\section{Methods} 

\subsection{HBAR Configurations}

\begin{figure} [htbp]
\includegraphics[width=1.0\linewidth]{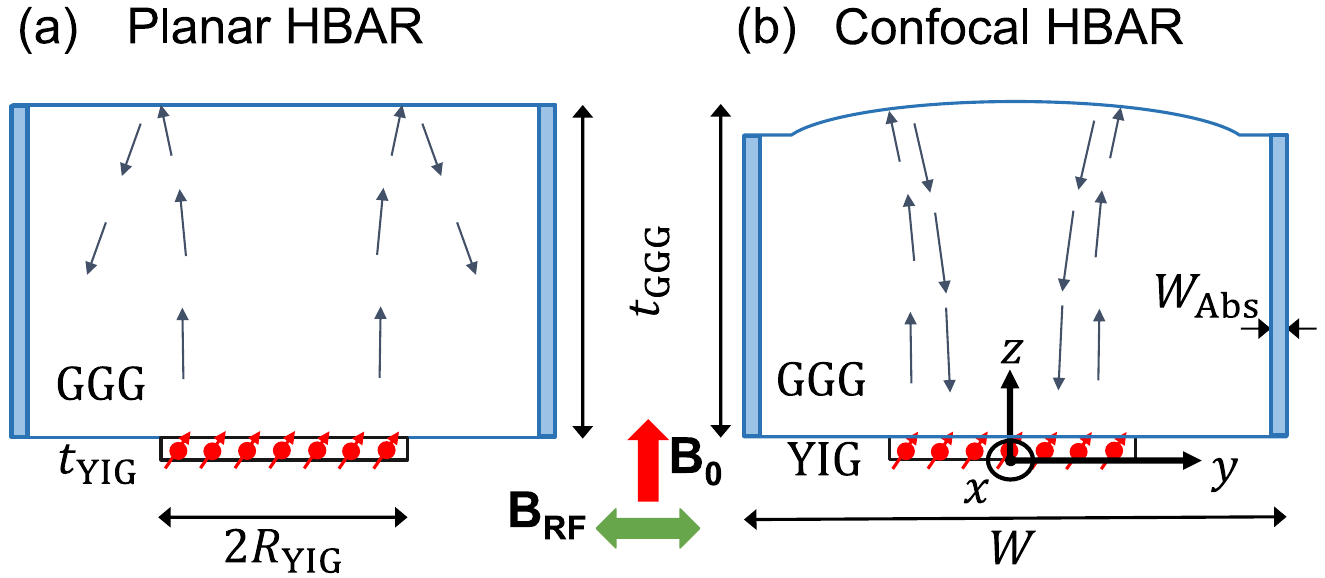}
\caption{{\bf Representative HBAR configurations:} (a) Planar HBAR structure consisting of a GGG thick film and a YIG thin film. (b) Confocal HBAR structure with a top dome and a planar bottom surface. $x$ axis is along the out-of-plane direction while $y$ and $z$ axes point along the lateral and the normal direction of the structures, respectively. Origin of the coordinate axes is at the center of the bottom YIG surface.}
\label{fig:HBARConfigs}
\end{figure}

Figure~\ref{fig:HBARConfigs} shows the two representative HBAR structures investigated in this work. The structure in Figure~\ref{fig:HBARConfigs}(a) consists of a thick GGG film joined with a YIG thin film at the bottom. We refer to this structure as the planar HBAR structure since it has planar top and bottom surfaces.  Figure~\ref{fig:HBARConfigs}(b) is a focusing HBAR structure that includes a GGG dome structure at the top and a planar bottom surface. We refer to this structure as the confocal HBAR (CHBAR) structure. The thickness of the GGG film for all configurations considered in this study, is $t_\text{GGG}=527\mu$m, as shown in Fig.~\ref{fig:HBARConfigs}. % This value refers to the length from the bottom surface of the GGG film to the mid point of the dome for the confocal HBAR, as shown. 
All YIG films are circular films or discs with thickness, $t_\text{YIG}=200$ nm, and radius of cross-section, $R_\text{YIG}$. The total HBAR device thickness is $t_\text{HBAR}=t_\text{GGG}+t_\text{YIG}=527.2\mu$m. We consider planar and confocal HBAR structures with several $R_\text{YIG}$'s ranging from 10$\mu$m to 200$\mu$m. The width $W$ of the structure is 1200$\mu$m unless otherwise mentioned. The radius of cross-section of the dome, $R_\text{cross}$, and its radius of curvature, $R_\text{curv}$, vary for different structures considered. We investigate 8 planar HBARs with varying $R_\text{YIG}$, 18 CHBARs with varying $R_\text{curv}$ and fixed $R_\text{cross}$, and 10 CHBARs with varying $R_\text{cross}$ and fixed $R_\text{curv}$.

\subsection{Magnon Modes}

The YIG thin film hosts magnons, generated by a static, $\textbf{B}_0$, and an oscillating RF magnetic field, $\textbf{B}_\textbf{RF}$. $\textbf{B}_0$ is along the $z$ axis, while $\textbf{B}_\textbf{RF}$ acts in the YIG plane. The magnetic fields, $\textbf{B}_\textbf{0}$ and $\textbf{B}_\textbf{RF}$, generate forward volume magnetostatic modes in the YIG film, that precess in the $x-y$ plane. The resulting dynamic magnetization is given by $\textbf{m}_{\text{YIG}} (x,y) = m_0 (x,y) (\text{cos}(\omega t)\textbf{i}+\text{sin}(\omega t)\textbf{j})$. Here, $m_0 (x,y)$ represents the magnon modeshape and $\omega$ is the frequency of precession. We practice the following convention throughout the article, bold-font symbols represent vector fields and corresponding normal-font symbols represent scalar values, respectively. We expect un-pinned spin waves at the top and bottom YIG film surfaces because of the following reason. The exchange interaction effects are expected to dominate at a thickness on the order of 200 nm and below. As a consequence, the pinning effect is expected to disappear below a critical width on the order of 200 nm~\cite{wang2019spin}. However, we expect full-pinning of the magnetic spin waves at the lateral boundaries of the YIG discs of radii ranging from 10 micrometers ($\mu$m) to 200$\mu$m. This is because the magnetic dipolar interaction effects dominate at length scales on the order of micrometers, over the exchange interactions~\cite{wang2019spin}. Taking these into account, we assume that $\textbf{m}_{\text{YIG}}$ is constant throughout the YIG film thickness. We describe the modeshape of the pinned spin waves, $m_0 (x,y)$, using truncated Bessel functions~\cite{guslienko2005boundary,kakazei2012probing}:
\begin{equation}
    m_0(x,y) = 
    \begin{cases}
        J_0 \left (\frac{R}{R_\text{YIG}}\zeta_j \right ),& \text{if } R\leq R_\text{YIG}\\
        0,              & \text{otherwise}
    \end{cases}
    \label{eq:MagnonMode}
\end{equation}
where radial coordinate, $R=\sqrt{x^2+y^2}$ and $J_0$ is the $0^\text{th}$ order bessel function. The zeroes of $\zeta_j$ with $j={0,1,2,...}$, correspond to the fundamental and higher-order magnon modes respectively. In this article, we consider the fundamental magnon mode or the Kittel mode, whose amplitude is represented by the function, $J_0(\frac{R}{R_\text{YIG}}\zeta_0)$. The modeshape is constant along the $z$-direction.

%The Kittel mode could also exist under different pinning conditions, which could lead to different magnon and phonon mode profiles discussed later in this article.
% if we increase thickness beyond 200 nm, we may see pinning along thickness, and laterally we may have to go much below 10$\mu$m for pinning effects to decrease, which is too small to consider for now.
%The Kittel mode could exist even if the magnetic spin waves are not fully pinned. However, exploration of such cases is left for future research.
%Such pinning conditions lead to different magnon and phonon mode profiles in the HBAR structures. We discuss the different case later in this article.

\subsection{Phonon Modes}

The forward volume magnon modes, $\textbf{m}_{\text{YIG}}(x,y)$, generate precessing shear deformations in the YIG region. The dynamic shear strain results in a circularly polarized chiral phonon traveling wave in the GGG region. The helicity of the chiral phonon is determined by the precession direction of the Kittel mode. The shear displacements can be expressed as $\textbf{u}_{\text{YIG}} (x,y) = u_0 (x,y) (\text{cos}(\omega t)\textbf{i}+\text{sin}(\omega t)\textbf{j})$, where $u_0 (x,y)$ is the phonon modeshape and $\omega$ is the precession frequency. In the following, we refer to the shear displacements, $\textbf{u}_{\text{YIG}} (x,y)$, as $\textbf{u}_0 (x,y)$. The traveling wave undergoes reflections at the top GGG surface and the reflected waves interfere with the forward traveling wave. When the forward and the reflected helical propagating waves interfere, they form rotating standing shear wave modes at specific frequencies. 
% They can be described as standing shear wave mode in a rotating coordinate system, rotating with frequency $\omega$. 
While the traveling chiral phonons are helical, the standing waves are not helical. The top GGG surface does not induce a $\pi$ phase shift to the reflected wave, unlike circularly polarized light reflecting off a mirror. As a result, we obtain standing shear wave modes with zero net helicity. In this article, we use (1) Fourier beam propagation and (2) HT eigenvalue problem approaches to analyze the shear phonon modes in the chosen HBAR configurations. 

\subsubsection{(1) Phonon Modeshape Analysis: Fourier Beam Propagation}

The Fourier beam propagation method (FBPM), also known as the angular spectrum method, predicts the field displacements or profiles of propagating waves~\cite{renninger2018bulk,matsushima2009band}. The advantage of FBPM lies in its simplicity and the ability to predict the field profiles at any target distances from the source without needing to calculate the behavior at intermediate distances. The FBPM has been widely used to analyze beam propagation in the field of optics~\cite{matsushima2009band}. The mathematical formulation of FBPM for acoustic waves 
% in anisotropic systems 
is well established~\cite{renninger2018bulk}. Recently, it has been used to study phonons in planar~\cite{chu2017quantum} and CHBAR structures~\cite{renninger2018bulk}, adapting an iterative method similar to the Fox-Li approach~\cite{Fox1961}. Here, we implement a reformulated iterative approach in which the propagation is calculated using projector and propagator operators. The reformulation allows us to achieve a seven-fold speed up of computation time. 
% The speed up happens if one wishes to use the standard FBPM which is presented in this section by re-formulating it using projectors and propagators. It doesn't happen if they follow adaptive algorithm that we ultimately use. 
Our approach is particularly advantageous for isotropic systems such as YIG and GGG. 

We assume an initial shear displacement field with modeshape
\begin{equation}
    u_0^{(1)}(x,y) = 
    \begin{cases}
        J_0 \left (\frac{R}{R_\text{YIG}}\zeta_0 \right ),& \text{if }R\leq R_\text{YIG}\\
        0.              & \text{otherwise}
    \end{cases}
    \label{eq:InputField}
\end{equation}
The iterative procedure begins with this initial input field and the superscript represents iteration index ($i=1$). The subscript in $ u^{(1)}_0(x,y)$ refers to the fact that the displacement is computed at $z=0$. Note that the lateral phonon modeshape is the same as the magnon Kittel modeshape, as shown in Eq.~\ref{eq:MagnonMode}. In FBPM, the input beams for initial and the subsequent iterations are decomposed into plane-waves. The field displacements at intermediate distances is then obtained by multiplying phase factors to the decomposed beam. The Fourier decomposition of the input beam is given by:
\begin{equation}
    \Tilde{\textbf{u}}^{(i)}_0(k_x,k_y) = \text{FFT}[\textbf{u}^{(i)}_0(x,y)].
    \label{eq:InputFFT}
\end{equation}
Here, $i$ is the iteration index and $\Tilde{\textbf{u}}^{(i)}_0$ is defined on a $N \times N$ $(k_x,k_y)$ grid which is the Fourier conjugate of the spatial $N \times N$ $(x,y)$ grid. We calculate the projections of $\Tilde{\textbf{u}}^{(i)}_0(k_x,k_y)$ along the different polarization directions, $m=1,2,3$, using the projector, $\textbf{P}_m$, and obtain the amplitudes, $A^{(i)}_m$, of the shear ($m=1,2$) and the longitudinal ($m=3$) modes: 
% \begin{subequations}
% \label{eq:PWAmplitudeCoeff} 
% \begin{align}
%         &A^{(i)}_m = \textbf{P}_m \cdot \Tilde{\textbf{u}}^{(i)}_0, \label{eq:Amplitudes}\\
%         \text{with } &\textbf{P}_m = \frac{\Sigma_{cyc} (-1)^{|\text{sgn}(i-m)|}\hat{d}_i(\hat{d}_m\cdot \hat{d}_i-\hat{d}_j\cdot \hat{d}_k)}{1+2\Pi_{cyc} (\hat{d}_i\cdot \hat{d}_j)-\Sigma_{cyc} (\hat{d}_i\cdot \hat{d}_j)^2}.
%         \label{eq:Projectors}%
% \end{align}
% \end{subequations}
\begin{subequations}
\label{eq:PWAmplitudeCoeff} 
\begin{align}
        &A^{(i)}_m = \textbf{P}_m \cdot \Tilde{\textbf{u}}^{(i)}_0, \label{eq:Amplitudes}\\
        \text{with } &\textbf{P}_m = \frac{\Sigma_{cyc} (-1)^{|\text{sgn}(j-m)|}\hat{d}_j(\hat{d}_m\cdot \hat{d}_j-\hat{d}_k\cdot \hat{d}_l)}{1+2\Pi_{cyc} (\hat{d}_j\cdot \hat{d}_k)-\Sigma_{cyc} (\hat{d}_j\cdot \hat{d}_k)^2}.
        \label{eq:Projectors}%
\end{align}
\end{subequations}
Here, $\hat{d}_m$ is the polarization vector for the plane-waves propagating along $(k_x,k_y,k_{z,m})$, $(j,k,l)$ refers to the Cartesian directions, $(j,k,l)= (1,2,3)$ and $\Sigma_{cyc}$ and $\Pi_{cyc}$ are cyclic sum and product operators, respectively, that cycle through variables $j$, $k$, and $l$. We use the amplitudes $A^{(i)}_m(k_x,k_y)$, to obtain the displacement field of the propagated beam, starting from the input beam, $u^{(i)}_0(x,y)$. For a wave originating in the YIG thin film, the propagation distance to reach the upper GGG surface of the HBAR structures is $t_\text{HBAR}$, as shown in Fig.~\ref{fig:HBARConfigs}. The displacement field of the propagated beam at the upper GGG surface is given by
\begin{subequations}
\label{eq:PWPropagator}
\begin{align}
        &\textbf{u}^{(i)}_{t_\text{HBAR}}(x,y) = \text{IFFT}[\Sigma_m A^{(i)}_m G_m],  \label{eq:IFFTProp}\\
        \text{with } &G_m(k_x,k_y) = \hat{d}_m(k_x,k_y) e^{ik_{z,m} (k_x,k_y) t_\text{HBAR}}. \label{eq:PWProp}
\end{align} 
\end{subequations}
Here, $G_m(k_x,k_y)$ is the propagator for plane-waves traveling along $(k_x,k_y,k_{z,m})$ with polarization $m$. Typically, $k_{z,m}$ can be derived from their respective slowness surfaces~\cite{renninger2018bulk} for each polarization and values of $(k_x,k_y)$. However, the functional relation can be simplified to $k_{z,m} = \sqrt{\frac{\omega^2}{v^2_m}-k^2_x-k^2_y}$, for isotropic dispersions. Here, $\omega$ is the frequency of the initial wave and $v_m$ is the velocity of phonons with polarization $m$. We use the simplified relationship and the material properties of YIG and GGG, shown in Table.~\ref{tab:YIGGGGProps}, to compute $k_{z,m}$ for the propagating waves in our isotropic YIG/GGG HBAR structures. Since the properties of YIG and GGG are similar, we use GGG values for both YIG and GGG, for simplicity. This approximation can be further justified by considering that our structures are composed of GGG thick films with ultrathin YIG films bonded to it. Note that we compute $k_{z,m}$, $A_m$ (Eq. \ref{eq:PWAmplitudeCoeff}), and $\textbf{u}_{t_\text{HBAR}}(x,y)$ (Eq.~\ref{eq:PWPropagator}) only for the first iteration, for a given $\omega$. This aspect results in the seven-fold computational speed-up mentioned earlier. 
\begin{table}[h]
\centering
\caption{Material properties of YIG and GGG.}
\begin{tabular}{lccc}
\hline
 & Young’s modulus & Poisson’s ratio & Density\\
 & E (Pa) & $\nu$ & $\rho$ (kg/m$^3$) \\
\hline
YIG~\cite{Sokolov2016} & 0.2 $\times$ 10$^{12}$ & 0.29 & 5170 \\
GGG~\cite{Vitko2015} & 0.222 $\times$ 10$^{12}$ & 0.28 & 7080 \\
\hline
\end{tabular}
\label{tab:YIGGGGProps}
\end{table}
We choose the longitudinal polarization $\hat{d}_3$ to be along the unit vector \textbf{k} and the shear polarizations, $\hat{d}_{1,2}$, to be two mutually perpendicular unit vectors perpendicular to \textbf{k}. For the isotropic systems investigated in this article, $d_m$'s are mutually perpendicular and $A_m$ reduces to $A_m = \hat{d}_m\cdot \Tilde{\textbf{u}}$ at each $(k_x,k_y)$. 

The propagated plane-waves are periodic in the direction transverse to the beam propagation direction. When the diffracted waves reach the transverse boundaries of the computational domain, they introduce undesired reflections. To avoid these reflections, one needs to consider a sufficiently wide computational domain that can contain the waves even after multiple reflections. However, such large domains can significantly increase computational costs. An alternative approach is to introduce absorbing boundary regions that completely attenuate any waves that enter these regions~\cite{chu2017quantum}. In this study, we implement absorbing boundaries of thickness $W_\text{Abs}=50\times\lambda$, where $\lambda$ is the wavelength of the input field. The blue shaded regions in Fig.~\ref{fig:HBARConfigs} show the absorbing boundaries. To simulate the effect of absorbing boundaries, we multiply $\textbf{u}^{(i)}_{t_\text{HBAR}}$ (Eq.~\ref{eq:PWPropagator}) with a reflection operator $R_{t_\text{HBAR}}$, defined as
\begin{equation}
    R_{t_\text{HBAR}}(x,y) = 
    \begin{cases}
        1,& \text{if } R\leq W_\text{eff}/2\\
        0.             & \text{otherwise}
    \end{cases}
    \label{eq:ReflectionGGG}
\end{equation}
$W_\text{eff}=W-2W_\text{Abs}$ is the effective width of the simulation window without the absorbing boundaries. We propagate the attenuated reflected wave further through a distance $t_{\text{HBAR}}$, We implement the beam propagation by following the approach outlined in Eqs.~\ref{eq:InputFFT} - \ref{eq:PWProp}, to complete a full round trip.
%We use  Eq.~\ref{eq:ReflectionYIG} in place of Eq.~\ref{eq:ReflectionGGG} at the end of the second half of the round trip. 
After every round trip, we multiply the resulting complex displacement field by an additional phase $R_{0,m}$ for each polarization $m$: 
\begin{equation}
    R_{0,m}(x,y) = 
    \begin{cases}
        e^{i2k_{z0,m} t_\text{YIG}},& \text{if } R_\text{YIG}\leq R\leq W_\text{eff}/2\\
        1,&\text{if }R \leq R_\text{YIG}\\
        0.              & \text{otherwise}
    \end{cases}
    \label{eq:ReflectionYIG}
\end{equation}
where $k_{z0,m} = k_{z,m}(k_x=0,k_y=0)$. The phase factor is introduced due to the finite width of the YIG film compared to the GGG width, $W$. However, it is worth mentioning that this approximation is more appropriate for low-diffraction cases. We used this for all cases considered to simplify the analysis. We use the resulting displacement field as the new input field for the next iteration. We repeat this process for $N$ round trips. At the end of $N$ round trips, we calculate the complex sum of the displacement fields, $\textbf{U}_0(x,y)=\Sigma^N_{i=1}\textbf{u}^{(i)}_0(x,y)$ at $z=0$. Using the interference sum, $\textbf{U}_0(x,y)$, we restart the iterative process with $\textbf{U}_0(x,y)$ as the initial beam of the next restart, $\textbf{u}^{(1)}_0(x,y) = \textbf{U}_0(x,y)$. Such a restart process using interference sum as an input ensures fast convergence to the desired mode. The other mode components are attenuated in the interference sum due to destructive interference. We continue this process until the varation of the modeshape is within a chosen tolerance. We obtain a converged standing shear wave with displacement field, $\textbf{u}_0(x,y)$, as a final outcome. 

In addition to the displacement profiles, we are interested in identifying the frequencies of the shear modes. These modes could couple with the Kittel magnon modes generated in the YIG thin-film (Eq.~\ref{eq:MagnonMode}), in a rotating coordinate system. The frequency overtones for plane-waves traveling along z-direction in the HBAR structures are expected to be $\omega_{m,n} =2\pi \times \frac{n v_m }{2t_{\text{HBAR}}}$, with $n = 1,2,3,...\infty$. Here, $t_{\text{HBAR}}$ is the thickness of the structure, velocity of wave is $v_m$ and $m$ represents polarization. The overtones are separated by $\frac{v_m }{2t_{\text{HBAR}}}$, known as the free spectral range (FSR). However, unlike plane-waves, the diffracting waves traveling in the $z$ direction do not have well-defined $k_{z,m}$'s leading to Gouy phase effects~\cite{feng2001physical}. The diffraction results in the shift of resonance frequencies from the monochromatic plane-wave overtones. To identify the resonance frequencies, we select beam of frequencies from a chosen frequency window, propagate the beam in the structure, and calculate the intensities of the interference sums ($I = \int_A \text{Re}[\textbf{U}_0 (x,y)]^2 dA$) at $z=0$ after $N$ round trips. The frequencies for which the intensities are maximum are the resonance frequencies of the shear waves in the HBAR structure. We focus on the shear modes with $m=2$, however, we could have as well chosen $m=1$ as the dominant shear polarization in the rotating coordinate system. We sweep through a frequency window of $\omega_0 \pm 5\times \text{FSR}$ using 50 steps. Here, the frequency of interest, $\omega_0=2\pi \times 9.825$ GHz, corresponds to the frequency of the 2960$^{th}$ overtone of a standing plane-wave in the HBAR structure. We choose this overtone to match with the structure investigated in a previous article~\cite{xu2021coherent}. We consider a YIG/GGG HBAR structure with $t_\text{YIG} = 200$ nm, $R_\text{YIG} = 200 \mu$m, $t_\text{GGG} = 527 \mu$m, and $L_{x,\text{GGG}} = L_{y,\text{GGG}} = 1200 \mu$m, for this analysis. 
\begin{figure} [htbp]
\includegraphics[width=1.0\linewidth]{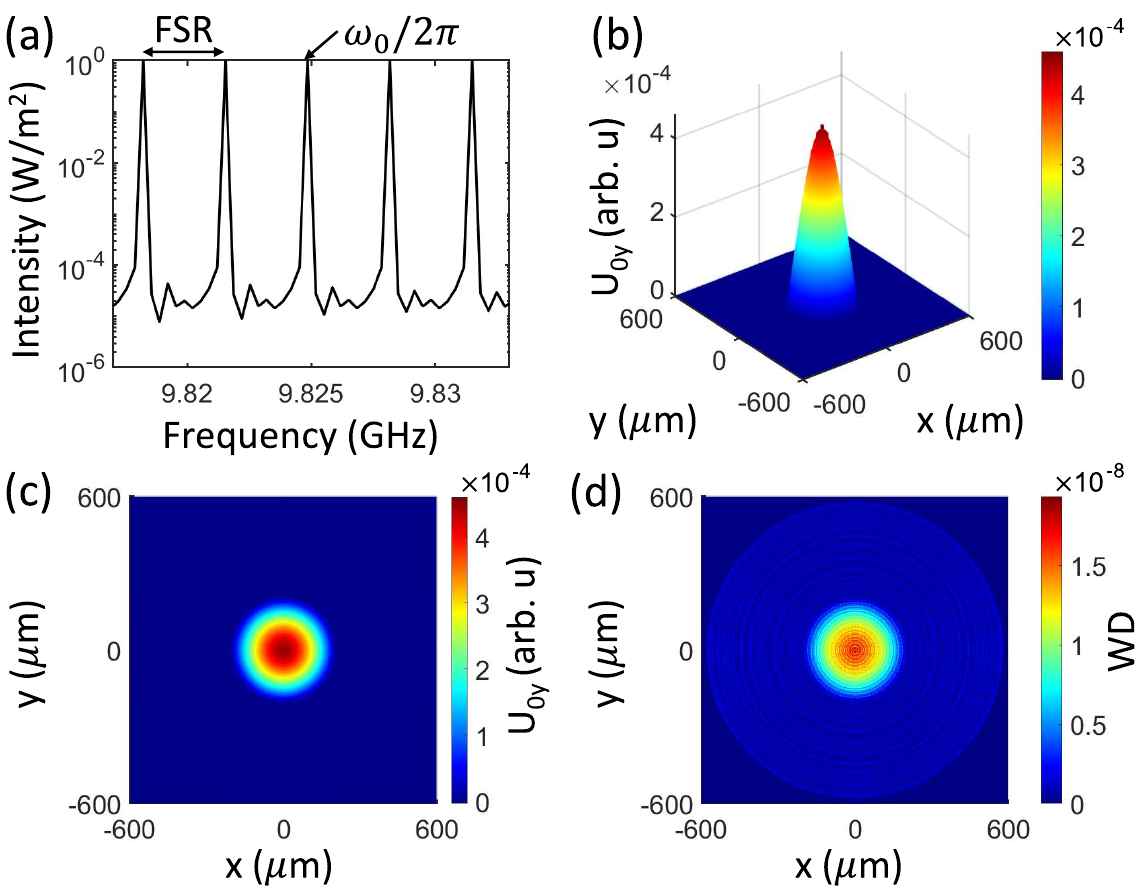}
\caption{{\bf Resonant modes in planar HBAR structures:} (a) Multimode phonons, represented by the overtones of the fundamental mode, and their free spectral range (FSR). (b) Isometric view of the y-component of $\textbf{U}_{0} (x,y)$, $\text{U}_{0y} (x,y)$. (c). Top view of $\text{U}_{0y} (x,y)$ showing localization effect caused by the YIG film. (d) Weighted deviation (WD) between mode profiles before first and second restarts.}
\label{fig:200umModesFSR}
\end{figure}

Figure~\ref{fig:200umModesFSR}(a) shows the intensities of different propagated beams with frequencies within the frequency window, a frequency window, $\omega_0 \pm 5\times \text{FSR}$. The central intensity peak corresponds to  $\omega_0=2\pi \times \frac{2960v_2}{2t_{\text{HBAR}}}$, the exact value of the $2960^{\text{th}}$ plane-wave overtone. The peaks form at resonance frequencies separated by FSR $=3.32$ MHz around the frequency, $\omega_0/2\pi \sim 9.825$ GHz, indicating the presence of multiple shear wave overtones. In Figs.~\ref{fig:200umModesFSR} (b) and (c), we show different views of the $y$-component of the normalized resonant modeshape, \text{Re}[\textbf{U}$_{0y} (x,y)$], at frequencies near $\omega_0$. We obtain these displacement fields after 800 round trips. This simulation included one restart after 400 round trips. We perform multiple tests and check that this number is large enough such that the Gouy phase effects are not significant on the interference sums. We obtain the normalized displacement fields before the 1st and the 2nd restarts, $\textbf{U}^1_0 (x,y)$ and $\textbf{U}^2_0 (x,y)$, respectively, and compute the weighted deviation (WD) between them: $|(\textbf{U}^1_0 (x,y)-\textbf{U}^2_0 (x,y))|/|\Sigma_{x,y}\textbf{U}^1_0 (x,y)|$.
% weighted deviation is usually defined as $|w(x,y)\times(\textbf{U}^1_0 (x,y)-\textbf{U}^2_0 (x,y))/\textbf{U}^1_0 (x,y)|/|\Sigma_{x,y}w(x,y)|$ usually defined as percentages, we get the above expression is we chose the weights to be $w(x,y)=\textbf{U}^1_0 (x,y)$. 
We monitor the WD between restarts to check for convergence. The color scale of Fig.~\ref{fig:200umModesFSR} (d) shows that the WD is much less than $10^{-7}$, indicating that the Figs.~\ref{fig:200umModesFSR} (b) and (c) represent displacement fields converged within sufficient numerical tolerance. Some fringes remain in the modeshapes that are possible artifacts of our numerical analysis. These are the high frequency components resulting from the sampling of the sharp phase change induced by the YIG film. They are significantly lower in values, however, could be further reduced by either using a low-pass filter or a smoother phase change factor than the function, $R_{0,m}(x,y)$ (Eq.~\ref{eq:ReflectionYIG}), used in this work. 

\begin{figure} [htbp]
\includegraphics[width=1.0\linewidth]{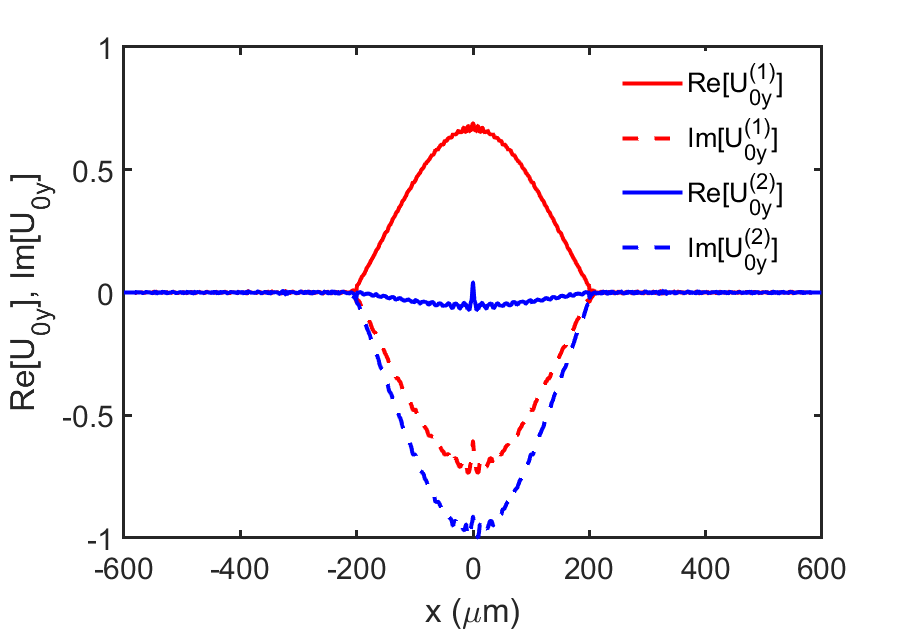}
\caption{{\bf Lateral mode profiles obtained with FBPM showing Gouy phase effects:} The real (solid) and imaginary (dashed) components of the y-component of the interference sum before the first $\text{U}^{(1)}_{0y}$ and second restarts $\text{U}^{(2)}_{0y}$ at $\omega_0$. The components deviate significantly between restarts.}. 
\label{fig:GouyPhaseProblem}
\end{figure}

To obtain the converged modes and identify the true resonant frequency near an intensity peak of interest, we further narrow the frequency sweeping range with finer sampling ($\sim 1$ kHz). We set the frequency to be $\omega^\text{HBAR}_0=\omega_0+\delta\omega$, with $\delta\omega=2\pi\times2.157$ kHz and restart the iterative process with Eq.~\ref{eq:InputField} as the input beam, to identify the true modal frequency near $\omega_0$. If such narrowing is not done, the modeshape can change significantly after each restart due to the Gouy phase effects the beam incurs due to diffraction~\cite{feng2001physical}. These effects results in the detuning of overtones with respect to the plane-wave overtones. Figure~\ref{fig:GouyPhaseProblem} shows the real and imaginary parts of the complex sums \textbf{U}$^{(1)}_{0y} (x,y)$ and \textbf{U}$^{(2)}_{0y} (x,y)$ computed before first and second restart (after 400 and 800 round trips), respectively. We calculate the complex sums at $\omega_0$ without performing the narrowed frequency sweeping discussed above. Both the real and imaginary parts of $\text{U}^{(1)}_{0y}$ and $\text{U}^{(2)}_{0y}$ show significant variations between the restarts. These results establish that the $\delta\omega$ correction is necessary to obtain the converged modes. 

However, the gradual narrowing of the frequency sweeping is a tedious process to identify the necessary fine sampling. We need to restart the iterative process multiple times especially for high-diffraction cases. The results shown in Fig.~\ref{fig:200umModesFSR}, represent propagating waves in a low-diffraction regime with Fresnel number, $N_F = 213$. However, this work also considers wave propagation in HBAR structures with $R_{\text{YIG}}$ as low as $10\mu$m corresponding to a very low Fresnel number, $N_F = 1.06$, indicating a high-diffraction regime. We investigate these structures to discuss how a reduced actuator lateral area affects the phonon modes and their lifetimes. The reduced area promises to increase the integration density of the HBAR devices towards high-density memory applications. The high-diffraction regime of these structures introduces significant Gouy phase effects on the propagating waves and results in detuning of frequencies. It becomes challenging to identify a resonant frequency by merely narrowing the frequency sweeping range. We implement an adaptive FBPM algorithm to circumvent this challenge. Note that we implement absorbing boundaries in this study to avoid undesired boundary reflections. The phonon modeshapes can be well predicted using this implementation. However, their lifetimes can be dependent on the shape and size of the boundaries. We leave the extensive investigation of the effect of boundary conditions on phonon lifetimes for a future investigation.
%Interesting insight: Large wavelength phonons may be sampled by a coarse grid but they may need a wider computational boundary because of their high-diffraction. On the other hand, small wavelength phonons need to be sampled using a fine grid, however one may do away with a smaller computational boundary due to their small diffraction. There may be a computational trade-off involved.

\subsubsection{Phonon Modeshape: Adaptive Fourier Beam Propagation}

We formulate the FBPM approach described above as an eigenvalue problem:
\begin{equation}
    \textbf{RT}\textbf{u}_\text{0}(x,y) =\Lambda\textbf{u}_\text{0}(x,y)
    \label{eq:EigValProblem}
\end{equation}
where \textbf{RT} is the round trip operator that includes all the operations described in the previous section (Eqs.~\ref{eq:InputFFT}-~\ref{eq:ReflectionYIG}) and $\Lambda$ is the eigenvalue corresponding to the eigenvector $\textbf{u}_{0}$. It is possible to solve the eigenvalue problem for 2D problems (e.g., if HBARs are represented by planar 2D structures), however, it cannot be directly solved for 3D structures of our interest. We develop an iterative method to obtain the resonant mode that satisfies Eq.~\ref{eq:EigValProblem} with a real $\Lambda$. $\Lambda$ is real for a standing wave mode. A complex $\Lambda$ applies a global phase to the input modeshape after a round trip, and the phase prevents the waves to interfere constructively over multiple round trips. In the adaptive FBPM (a-FBPM) algorithm, we iteratively adjust the frequencies based on the phase difference incurred over a round trip to arrive at the desired standing wave modes. We outline the steps of the iterative method below. 

\begin{enumerate}
    
\item Start with a input beam, $\textbf{u}^{(i)}_\text{0}(x,y)$, for the $i^{th}$ round trip. The initial input beam $(i=1)$ has wavelength, $\lambda^{(1)}$ $=\frac{2t}{n}$ and frequency $\omega^{(1)} =2\pi \times \frac{n v_2 }{2t}$. Here, $n$ is the overtone number, $t=t_{\text{HBAR}}$ and $v_2$ is the velocity of shear phonons. The wavelength and frequency are estimated based on the $n^{\textbf{{th}}}$ overtone for an open-ended column. The inital modeshape is chosen to be a truncated Bessel function as shown in Eq.~\ref{eq:InputField}. 

\item Calculate $\textbf{u}^{(i+1)}_\text{0}(x,y)=\textbf{RT}\textbf{u}^{(i)}_\text{0}(x,y)$.

\item Estimate the real eigenvalue from, $\Lambda^{(i)} = \sqrt{\frac{I^{(i+1)}}{I^{(i)}}}$, where $I^{(i)}=\int |\textbf{u}^{(i)}_\text{0}|^2 dxdy$ and $I^{(i+1)}=\int |\textbf{u}^{(i+1)}_\text{0}|^2dxdy$.

\item Calculate the residual, $\text{R}^{(i)} = ||(\textbf{u}^{(i+1)}_\text{0}(x,y) -\Lambda^{(i)}\textbf{u}^{(i)}_\text{0}(x,y))||/||\Lambda^{(i)}\textbf{u}^{(i)}_\text{0}(x,y)||$.

\item If R$^{(i)}$ $\leq \text{tolerance (tol)}$, end simulation and output final eigenvalue and modeshape of resonant modes, else continue to the next step. We choose $\text{tol}=10^{-6}$ for all a-FBPM calculations of this study.

\item If R$^{(i)} > \text{tol}$, estimate the global phase factor introduced by the round trip operation $\textbf{RT}$, $\theta^{(i)} = \text{Arg}(\textbf{u}^{(i)}_\text{0}(0,0))-\text{Arg}(\textbf{u}^{(i+1)}_\text{0}(0,0))$.

\item Set $\omega^{(i+1)}= \omega^{(i)}+\frac{v_2\theta^{(i)}}{2t_{\text{HBAR}}}$ and $\lambda^{(i+1)}=2\pi n v_2/\omega^{(i+1)}$. The steps 6 and 7 makes the algorithm adaptive.

\item Repeat the process until step 5 is satisfied or the maximum number of $N$ round trips is reached, at which point we set $\textbf{u}^{(1)}_0(x,y) = \textbf{U}_0(x,y)=\Sigma^N_{n=1}\textbf{u}^{(i)}_0(x,y)$, and restart the iterative procedure. 

\end{enumerate}

\begin{figure} [htbp]
\includegraphics[width=1.0\linewidth]{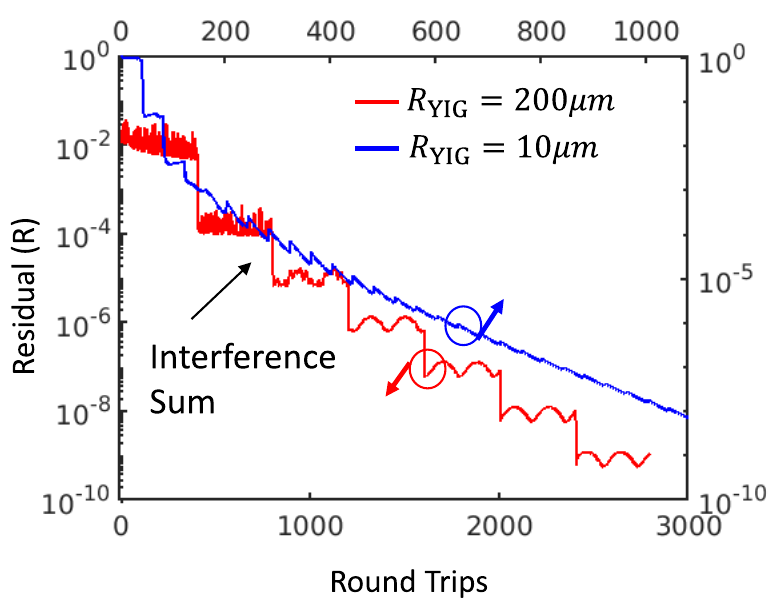}
\caption{{\bf Obtaining resonant modes using the adaptive FBPM algorithm:} Residual (R) of modeshapes decays as a function of number of round trips. Results are shown for two planar HBAR structures with $R_\text{YIG}=$ 200 $\mu$m (red, left and bottom axes) and 10 $\mu$m (blue, right and top axes). Circles with arrows point to corresponding axes for each plot.} 
\label{fig:200umModesNIConv}
\end{figure}

Figure~\ref{fig:200umModesNIConv} illustrates the effectiveness of the a-FBPM algorithm in predicting the resonant modes of the HBAR structures. We show the results for two planar HBAR structures with $R_\text{YIG}=$ 200 $\mu$m (red) and 10 $\mu$m (blue), respectively. The residual, R, decays as a function of the number of steps, as we approach the resonant mode. The step-like features of the residual decay occur after every $NR$ round trips when we compute the interference sums and use it as input for the next restart of the iterative procedure. The jumps occur because some errors get canceled due to destructive interference when an interference sum is computed. In some cases, R can have a momentary rise due to accumulation of errors from previous round trips, however, the overall trend continues to decrease ultimately reaching convergence. We calculate the complex interference sum after every $NR$ round trips. We choose $NR = 400$ and 40 for the HBARs with $R_\text{YIG}=$ 200 $\mu$m and 10 $\mu$m, respectively. We choose a smaller number of round trips, $NR$, for the 10 $\mu$m case, since the HBAR with $R_\text{YIG}=$ 10 $\mu$m represents a high-diffraction structure and the modeshape decays rapidly for this case. As shown in Fig.~\ref{fig:200umModesNIConv}, it takes a total of $\sim$650 and $\sim$1200 round trips (including restarts) to obtain converged resonant modes for the 10 $\mu$m and the 200 $\mu$m case, respectively. We continue the iterative process till R $<10^{-8}$ to show stability of the resonant mode even after tolerance is reached. 

\subsubsection{(2) Phonon Modeshape Analysis: Hankel Transform Eigenvalue Problem}

Our a-FBPM approach avoids the frequency sweeping process and mostly overcomes the slow convergence issues of the standard FBPM approach. However, it is still challenging to obtain converged phonon modes for the confocal HBAR structures of interest, using the a-FBPM approach. Here, we discuss an alternate method that uses the Hankel transform (HT) approach, the axi-symmetric equivalent of the Fourier transform method. The HT approach has been widely used in the field of optics, e.g., Fabry-Perot cavities~\cite{poplavskiy2018fundamental}. However, it has not been applied for acoustics problems, to the best of our knowledge. The HT method allows us to obtain converged phonon modes with significantly reduced computational cost. The approach leverages the axi-symmetry of the problem to reduce the 3D problem to a 2D one. For example, the 2D problem can be obtained by representing HBARs as planar 2D structures. The axi-symmetry conditions are applicable for the YIG/GGG HBAR structures due to their isotropic material properties and the circular cross-section of the YIG film. Note that the scalar field describing the dominant shear-component of $\textbf{u}_0$ is expected to obey axi-symmetry, however, the vector field of the shear acoustic phonon modes does not obey axi-symmetry. Here, we provide a brief description of the HT method. We encourage the interested reader to find a detailed description of the method elsewhere~\cite{poplavskiy2018fundamental,vinet1993matrix}. 

We cast the HBAR structures shown in Fig.~\ref{fig:HBARConfigs} into axi-symmetric representations about the $z$-axis. We transform the $x-y$ coordinates to the radial coordinate $r$, with limit $0\leq r \leq W/2$. We write the Hankel eigenvalue problem as $\textbf{RT}_\text{hk}u(r)=\Lambda u(r)$, where $u(r)$ represents the axisymmetric scalar phonon modeshape, $\Lambda$ is the corresponding eigenvalue and $\textbf{RT}_\text{hk}$ is the round trip operator. Since the dimensionality of the problem is reduced from 3D to 2D, the Hankel eigenvalue problem can be solved to obtain the phonon modes directly. We discretize $r$ as
\begin{equation}
    r_j = W/2\left(\frac{\zeta_j}{\zeta_N}\right) 
    \label{eq:HKdiscretizaiton}
\end{equation}
where $j=1,2,3,..,N$ and $\zeta_j$ ($j=1,2,3,..,N$) are the roots of the $J_1$ Bessel function. The round trip operator for the HT approach $\textbf{RT}_\text{hk}$ is given by
\begin{equation}
\textbf{RT}_\text{hk}=\textbf{R}_0\textbf{P}\textbf{R}_{t_\text{HBAR}}\textbf{P} 
    \label{eq:HKRoundTrip}
\end{equation}
where $\textbf{P}$, the $N\times N$ propagator matrix, is defined as:
\begin{subequations}
\label{eq:HKPropagator}
\begin{align}
        \textbf{P}&=(H^+)^{-1} \Tilde{G} H^+,\\
         H^+_{ij}&=\frac{W^2}{2\zeta_N^2}\frac{J_0(\zeta_j\zeta_j/\zeta_N)}{\zeta_N^2J^2_0(\zeta_j)},\\
          \Tilde{G}_{ij} &= \text{exp}\left( -i\frac{2\zeta^2_j}{W^2}\right)\delta_{jj}.\label{eq:HKHBARreflection}
\end{align}
\end{subequations}
Here, $H^+$ is the HT whose matrix inverse is taken to be an approximation of the inverse HT, and $\Tilde{G}$ is the Green's function obtained from the Fourier transform of the Fresnel propagator in the paraxial approximation. $\textbf{R}_{t_\text{HBAR}}$, and $\textbf{R}_0$ are the $N\times N$ diagonal reflection matrix operators at $z=t_\text{HBAR}$ and $z=0$, respectively and the diagonal elements are defined as
\begin{subequations}
\label{eq:HKReflection}
\begin{align}
        \textbf{R}_{0,jj}&=R_{0,2}(r_j),\\
         \textbf{R}_{t_\text{HBAR},jj}&=R_{t_\text{HBAR},2}(r_j).
\end{align}
\end{subequations}
This method, unlike the FBPM/a-FBPM approaches, does not require a Fox-Li-like iterative process and can also predict higher-order phonon modes. Although this approach has a limited applicability due to its axi-symmetric constraints, it makes it possible to obtain the phonon modes and lifetimes of HBAR structures at a reduced cost. The expedited analysis allowed us to analyze a large class of HBAR structures, as we show in the Results and Discussion section. 

% Therefore, at this point, it is difficult to envision further applications using this approach. Perhaps, an axisymmetric metasurface, such as a Fresnel reflector. But that's about it. Coming up with a few applications will make the inclusion of this method more valuable. It also seems to give inconsistent results with FBPM for high-diffraction cases. But, both are numerical methods with roots in Fourier transform, with the FBPM considering the other (x-z) displacement components as well. 
% $\tau$ and $g_\text{mb}$ predictions match very well with that of FBPM's for planar HBAR structures with varying YIG radii (although their modeshapes start differing from each other a bit as we go towards high-diffraction cases). 
% After which I presented the Hankel predictions for CHBAR structures, because I couldn't get FBPM's predictions to converge for all $R_\text{curv}$ and $R_\text{cross}$

\subsubsection{Diffraction-Limited Phonon Lifetime}

We estimate the diffraction-limited phonon lifetimes using three methods: ({\bf A}) Eigenvalue method, ({\bf B}) Exponential curve fitting method, and ({\bf C}) Clipping method.
% \blue{The more I think about it, the clipping method doesn't really seem applicable in our case. We did mention its limited applicability somewhere. Question is should we include this at all?}

\begin{figure} [htbp]
\includegraphics[width=1.0\linewidth]{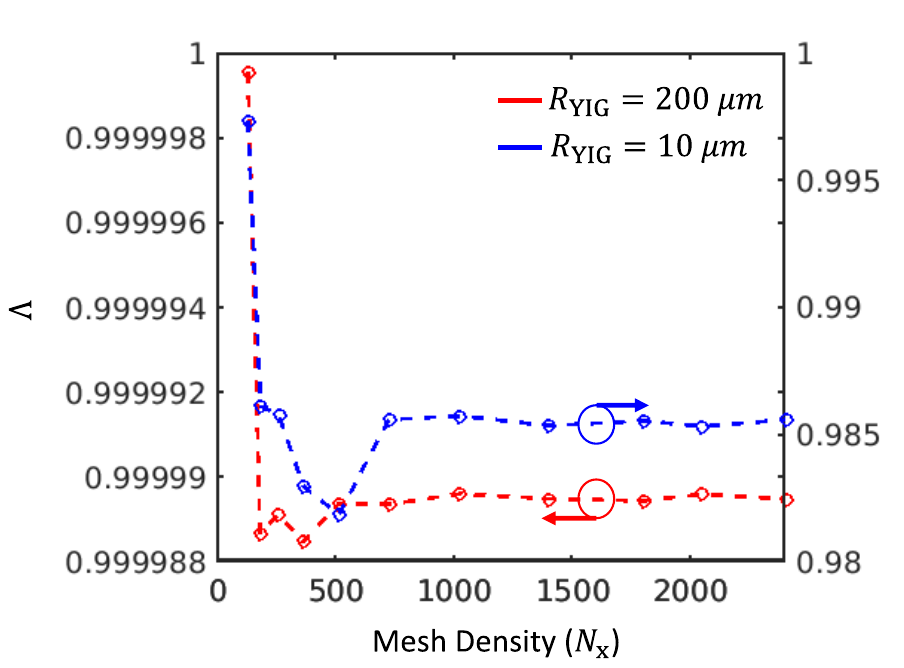}
\caption{{\bf Eigenvalues of two planar HBAR structures with $R_\text{YIG}=200\mu$m (red) and $10\mu$m (blue):} Variation of $\Lambda$ for meshes with density ranging from $128\times128$ to $2400\times2400$. The width $W$ is kept fixed at 1200$\mu$m. The circles with arrows point to the axes corresponding to each plot.} 
\label{fig:LambdaNConv}
\end{figure}

({\bf A}) Eigenvalue method: In this method, we obtain the eigenvalues using the adaptive FBPM simulation and use them to estimate the phonon lifetime. The elastic energy contained in a acoustic beam with modeshape $\textbf{u}$ is given by $E\propto \int |\textbf{u}|^2 dxdy$. 
% We do not consider the $z$-dependence of \textbf{u} for the energy calculation, 
We only integrate over the $dxdy$ element since the modeshape largely remains unaltered throughout the thickness. We assume that the elastic energy of the acoustic beam decays exponentially with the propagation time. For a cavity phonon mode with a total initial elastic energy $E_{tot}$, we represent the elastic energy left in the cavity after one round trip as $E_{in}=E_{tot}e^{-t_\text{RT}/\tau}$. Here, $t_\text{RT} (=2t_\text{HBAR}/v_2)$ is the time taken by shear waves to complete a round trip and $\tau$ is the lifetime of the phonon mode. On the other hand, the ratio of $E_{in}$ to $E_{tot}$ is given by $E_{in}/E_{tot} = \Lambda^2$. Here, $\Lambda$ is the real eigenvalue obtained using the a-FBPM simulations. Using this relation, we can obtain the lifetime of the phonon mode as
\begin{equation}
    \tau = \frac{-2t_\text{HBAR}}{v_2\text{ln}(\Lambda^2)}.
    \label{eq:LifetimeEigVal}
\end{equation}
Since these computations are performed on a finite $x-y$ mesh grid, $\Lambda$, and consequently, $\tau$ can be sensitive to the mesh density. Hence, it is important to select an optimal grid to obtain the converged values of $\Lambda$. In Fig.~\ref{fig:LambdaNConv}, we show the variation of $\Lambda$ with mesh density $N_x$ ($=N_y$) for both the low-diffraction ($R_\text{YIG}=200\mu$m) and the high-diffraction ($R_\text{YIG}=10\mu$m) cases. We find that the variation of $\Lambda$ is much less than 1\% when $N_x \geq 1024$. Consequentially, we choose $N_x=N_y=1024$ to compute phonon modes and lifetimes for all cases considered here. 

\begin{figure} [htbp]
\includegraphics[width=1.0\linewidth]{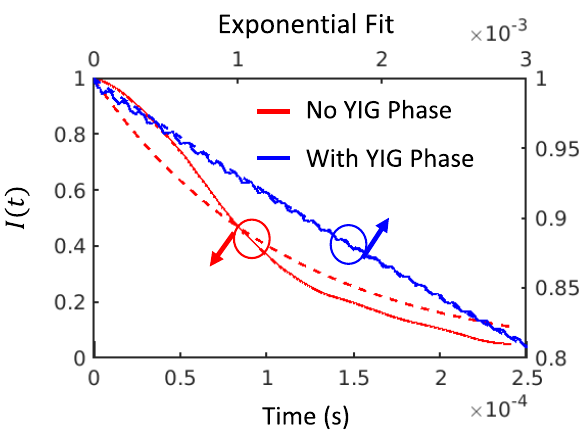}
\caption{{\bf Decay of the overlap function, $I(t)$, for propagating beams in a HBAR structure with $R_\text{YIG}=200\mu$m}. Solid lines indicate $I(t)$ while dashed lines indicate their respective exponential fits. The finite width of the YIG film induces a localizing effect on the beam, modeled with a phase factor. $I(t)$ decays at a much slower rate when the phase effect is included (blue, top and right axes) compared to the no-phase-effect case (red, bottom and left axes).}
\label{fig:200umQvsSvsAF}
\end{figure}

({\bf B}) Exponential curve fitting method: In this method, we calculate the phonon mode lifetime by evaluating the diffraction loss of the beam as it travels in the HBAR structure. We estimate the diffraction loss from the overlap of the propagated mode profile with the initial input profile, after each round trip. We estimate the overlap using the following function:
\begin{equation}
    I(t) = \frac{|\left<\textbf{u}_0(t)|\textbf{u}_0(0)\right>|^2}{|\left<\textbf{u}_0(0)|\textbf{u}_0(0)\right>|^2}.
    \label{eq:ExponentialFit}
\end{equation}
Here, $t$ is an integral multiple of the time taken for each round trip, $t_\text{RT}$. We allow the initial beam to travel for multiple round trips until $I(t)<0.1$ is reached or the time is $t>3$ ms, whichever is reached first. In Fig.~\ref{fig:200umQvsSvsAF}, we show the decay of the overlap function, $I(t)$, for beams traveling in a HBAR structure with $R_\text{YIG}=200\mu$m. The finite width of the YIG film (transducer) induces a localizing effect on the propagating beam. The localizing effect can be modeled by introducing a phase factor. The blue and the red lines shown in Fig.~\ref{fig:200umQvsSvsAF} correspond to the two cases when we obtained the overlap function with or without the consideration of the phase factor, respectively. The two different decays highlight the effect of the finite width of the transducer on the propagating beam. We find that $I(t)$ decays at a much slower rate when the YIG-induced phase factor is considered, compared to the case without the phase factor. We fit the two decays with exponential functions and obtain the phonon lifetimes as $\tau^{wophase}=0.1$ ms and $\tau^{wphase}=14.1$ ms, respectively. The remarkable two-orders of magnitude difference between these two predicted lifetimes highlights the importance of including the transducer or actuator phase effects for such an analysis. Subsequently, we included the phase effects in all our analyses to obtain the phonon lifetimes. We find that the lifetime predicted with the phase effects, $\tau^{wphase}=14.1$ ms, closely matches with $\tau$ of $15.5$ ms, predicted using the eigenvalue method discussed earlier. A past study used the exponential curve fitting approach to estimate the lifetime of the longitudinal HBAR phonons in an AlN/Sapphire structure ~\cite{chu2017quantum}. However, they treated the phonons as superpositions of Bessel functions 
% in a large simulation window 
and did not account for the localization effects induced by the AlN transducer. 

({\bf C}) Clipping method: In this method, we obtain the phonon lifetime using a similar expression as used for the eigenvalue method, Eq.~\ref{eq:LifetimeEigVal}: $\tau = \frac{-2t_\text{HBAR}}{v_2\text{ln}(E_{in}/E_{tot})}$. Here, $E_{tot}$ is the total initial elastic energy for a cavity phonon mode and $E_{in}$ is the elastic energy left in the cavity after one round trip. We calculate $E_{in}$, contained in a volume, $V_{in}$, of a cylinder with radius of the YIG disc and the thickness of the HBAR structure, $E_{in}\propto \int_{V_{in}} |\textbf{u}|^2 dxdy$. We consider a converged modeshape, obtained with a-FBPM or HT method. This method results in the lowest estimate of the phonon lifetimes since it assumes that all energy outside the cylinder of interest is lost after a round trip. Our approach is similar to the clipping method used for Fabry-Perot cavities~\cite{poplavskiy2018fundamental,vinet1993matrix} and HBAR resonators, excited opto-mechanically using photon beams~\cite{kharel2018ultra}. In these optical systems, the clipping method is more applicable since the input photon beam spans the entire $x-y$ plane and is clipped by the localizing finite cross-section mirrors or domes. In our system, the input acoustic beam is already laterally confined so this method may have limited applicability. 
% In fabry-perot cavities with finite mirrors, a gaussian input beam which spans the entire XY space will be clipped due to the finiteness of the mirrors with the additional beam wavefronts escaping into space. In our case, the input beam is already a finite beam in XY space, so clipping doesn't directly relate to the way it was originally used.
%Additionally, the planar surfaces besides the dome/disc also act as acoustic mirrors, so they don't necessarily clip like FP cavity. The losses here should come from boundary leakage (anchors/connections, lossy materials, BAW-SAW coupling), interfaces, surfaces, and ultimately material losses 
We still include the discussion here as a consistency check since the method results in the lower bound for the predicted lifetimes. 

\subsection{Magnon-Phonon Coupling}

The magnon-phonon coupling strength, $g_{mb}$, is given by
\begin{equation}
    g_{mb} = \frac{B}{\sqrt{A_\nu Q_\eta}}\left|\int_{V_\text{YIG}} \left[\frac{du_x^*}{dz}m_x+\frac{du_y^*}{dz}m_y\right] d\textbf{r}\right|.
    \label{eq:MagnonPhonon}
\end{equation}
Here, $B=7\times10^5$ J/m$^3$, is the magnetoelastic constant of YIG, $(m_x, m_y)$ are the $x$ and $y$ components of the magnetization, $\textbf{m}=\textbf{m}_\text{YIG}$, $(u_x^*, u_y^*)$ and $(\frac{du_x^*}{dz},\frac{du_y^*}{dz})$ represent the complex conjugates of the $x$ and $y$ components of displacement \textbf{u} and their corresponding shear strains, respectively. The integration domain is the volume of the YIG film, $V_\text{YIG}$, since the magnon modes reside in YIG. The term $\sqrt{A_\nu Q_\eta}$ corresponds to the magnon and phonon normalizing constants. We normalize the magnon modes as 
\begin{equation}
    \frac{M_s}{\gamma}\int_{V_\text{YIG}} \textbf{m}_\nu^*(\textbf{r})\cdot(\textbf{k} \times \textbf{m}_{\nu'}(\textbf{r})) d\textbf{r} = -iA_\nu\delta_{\nu,\nu'}.
    \label{eq:MagnonNormal}
\end{equation}
Here, $\mu_0 M_s=0.175$T is the saturation magnetization, $\gamma=2\pi\times28.5$ GHz/T is the gyromagnetic ratio, \textbf{k} is the unit vector along the z axis, $\nu$ is the magnon mode index, and $A_\nu$ is the magnon normalization constant, respectively. Similarly, we normalize the HBAR phonon modes as
 \begin{equation}
    2\omega_\eta \rho \int_{V_\text{HBAR}} \textbf{u}_\eta^*(\textbf{r})\cdot \textbf{u}_{\eta'}(\textbf{r}) d\textbf{r} = Q_\eta\delta_{\eta,\eta'}.
    \label{eq:PhononNormal}
\end{equation}
Here, $\rho=7080$ kg/m$^3$ is the GGG material density, $\eta$ is the phonon mode index, and $\omega_\eta$ is the corresponding phonon frequency, respectively. $V_\text{HBAR}$ is the volume of the HBAR structure. $Q_\eta$ is the phonon normalization constant and is of the same dimensionality as $A_\nu$. 

%Note that one could normalize the strain and magnetostatic energies to be $\hbar\omega$ in Eqs.~\ref{eq:MagnonNormal} and ~\ref{eq:PhononNormal} 
Note that if we assume zero diffraction, the acoustic energy is fully contained in the volume $V_{in}$, the volume of the cylinder with radius of the YIG disc and the thickness of the HBAR structure. Such a scenario results in a complete overlap of lateral mode profiles of the magnon and the phonon modes. We obtain the zero diffraction limit of the magnon-phonon coupling strength to be $g^0_{mb}/{2\pi} = 1.13$ MHz. This value is independent of the width of the YIG film. Our result compares well with a previous experimental result of $1$ MHz~\cite{an2020coherent}, and is slightly higher than another experimental result of $0.75$ MHz~\cite{xu2021coherent}. However, we consider the regime where diffraction causes the spread of acoustic energy outside $V_{in}$, which can impact both $\tau$ and $g_{mb}$ of phonon modes in HBAR structures. We present the diffraction-limited HBAR phonon lifetimes and the magnon-phonon coupling strengths in the Results and Discussion section.

\section{Results and Discussion} 

We investigate hybrid magnonic HBAR structures that can support the development of high-density phononic quantum memories. We can include a greater number of transduction components in a single chip of given dimension by reducing the radius of the YIG film. However, the reduced aperture increases the diffraction of the acoustic waves propagating into the GGG region. Here, we discuss the diffraction-limited phonon lifetimes and the magnon-phonon coupling in the HBAR structures.

\begin{figure*} [htbp]
\includegraphics[width=1.0\linewidth]{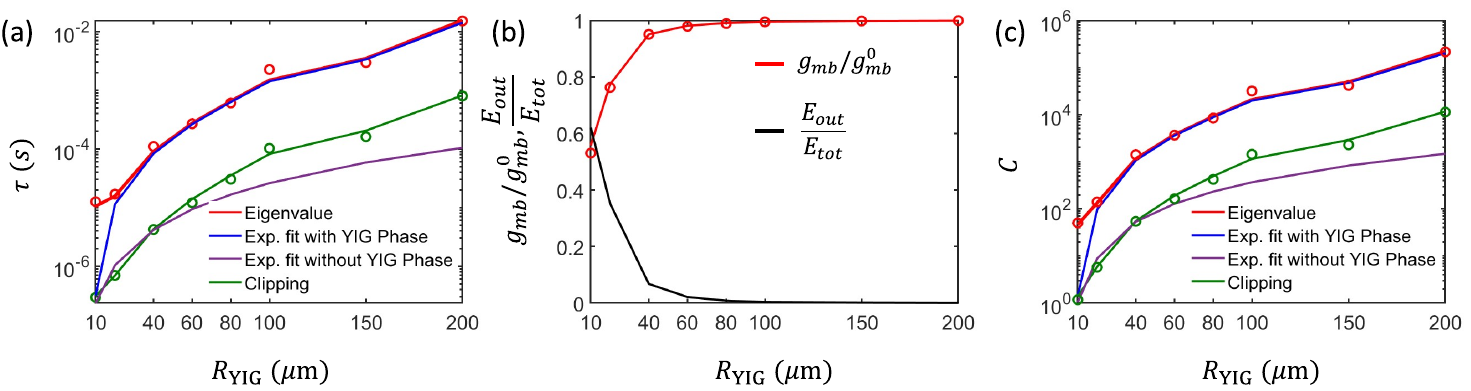}
\caption{{\bf Diffraction-limited properties of phonon modes of planar HBAR structures:} (a) Phonon lifetimes, $\tau$, calculated from the eigenvalue (red), exponential fitting (blue, purple), and clipping methods (green). (b) Ratio between magnon-phonon coupling in HBAR structures in various diffraction regimes and that in the zero-diffraction limit, $g_{mb}/g^0_{mb}$ (red). Reduction of $g_{mb}$ as we decrease $R_{\text{YIG}}$, is connected to the spread of modeshape in the high-diffraction regime. Increasing amount of acoustic energy spreads into the GGG region, causing reduced overlap between phonon and magnon modes in the YIG region. The ratio of acoustic energy spreading out to the total energy, $E_{out}/E_{tot}$ (black), increases in the high-diffraction regime. (c) Magnon-phonon cooperativity $C$. $R_\text{YIG}$ is varied keeping all other geometric factors fixed. Solid lines correspond to a-FBPM predictions, while the circles of the same color correspond to the respective HT predictions.}
\label{fig:GmbandTauvsRad}
\end{figure*}

\subsubsection{Diffraction-Limited Phonon Lifetime}

Figure~\ref{fig:GmbandTauvsRad} (a) shows the variation of phonon lifetimes in planar HBAR structures as we decrease the YIG radius. We compute the lifetimes using the a-FBPM approach and following the eigenvalue method (red line), the exponential fitting (blue and purple lines), and the clipping methods (green line), respectively. In addition to the a-FBPM predictions, we show predictions from HT method in Fig.~\ref{fig:GmbandTauvsRad} using circles, with colors matching to their corresponding a-FBPM predictions. Note that we use $|\Lambda|$ instead of $\Lambda$ in Eq.~\ref{eq:LifetimeEigVal} to calculate $\tau$ using the HT method. All methods predict that the lifetimes decrease with decreasing $R_\text{YIG}$ due to increased diffraction, as expected. The eigenvalue method results in the highest lifetime estimates, among the three methods considered. We obtain the highest lifetime to be $\tau=15.5$ms ($Q= \omega\tau=9.57\times10^8$) for HBAR structure with $R_\text{YIG}=200\mu$m. The lowest lifetime predicted by the eigenvalue method, is $\tau=10.5\mu$s ($Q=6.48\times10^5$), corresponding to HBAR structure with $R_\text{YIG}=10\mu$m, respectively. These results show that the lifetimes are reduced by more than three orders of magnitude in the high-diffraction regime. The lifetimes predicted by the eigenvalue method (red) closely match with those predicted by the exponential fitting method (blue), for the low-to-medium diffraction cases, when the phase effects are included. This result can be explained through the following argument. The input beam can be thought of as a superposition of the eigenmodes of the HBAR cavity. Once the initial input beam undergoes multiple round trips, only the fundamental mode is likely to survive, while the rest of the mode components decay at a faster rate. When we operate at the overtone frequency of one of the fundamental modes, the fundamental mode becomes the dominant mode. It can be argued that this dominant mode is the same as the fundamental eigenmode found by solving the eigenvalue equation, Eq.~\ref{eq:EigValProblem}. Therefore, the decay of the overlap function of the input field (exponential fitting method) is expected to have a similar character to that of the decay of the dominant eigenmode (eigenvalue method). As a result, we obtain similar lifetimes using the two methods. 

However, in the high diffraction case ($R_\text{YIG}=10\mu$m), we observe 35-fold reduced lifetimes predicted by the exponential fit compared to that of the eigenvalue method. Due to high-diffraction, significant amount of the energy of the input acoustic beam spreads out laterally during the first few round trips, resulting in a rapid initial decay of $I(t)$. In the exponential fit method, we obtain the lifetime from the exponential fit of $I(t)$, starting at $t=0$. The loss of acoustic overlap during the initial transient phase results in lower $\tau$ predictions. We expect that the two lifetime predictions will be closer if the exponential fit is obtained after a steady state is reached. Note that we obtain the exponential fitting predictions (blue) by including the localizing phase effects due to the YIG film. We also show the lifetimes predicted by this method, when we do not consider the phase effects due to the YIG film (purple). We find that the $\tau^{wophase}$'s are consistently lower than the $\tau^{wphase}$'s. The difference between the two predictions decreases monotonously from 135 to 1.35 fold, as we decrease $R_\text{YIG}$ from $=200\mu$m to $=10\mu$m, respectively. This is expected since the localizing effect of the YIG film diminishes as we approach $R_\text{YIG} \rightarrow 0$. Past studies often ignored the effect of aperture width on the HBAR phonons, however, our results (Fig.~\ref{fig:200umQvsSvsAF} and Fig.~\ref{fig:GmbandTauvsRad}) establish that such effects needs to be considered to make reliable predictions.  

Figure~\ref{fig:GmbandTauvsRad} (a) also shows the lifetime predictions from the clipping method (green), which assumes that the energy outside the cylindrical volume of interest, $V_{in}$, is completely lost after a round trip. We find the lifetimes to be more than an order of magnitude lower than those predicted from the eigenvalue method, for all cases considered. We obtain the lowest lifetime in the high-diffraction regime to be $\tau = 0.311\mu$s ($Q = 1.92\times10^4$). Interestingly, this prediction is still marginally greater than the experimentally observed values of 0.25$\mu$s and $1.45\times10^4$, respectively~\cite{xu2021coherent}, for a device operating in the low-diffraction regime. The reason for this discrepancy is that the experimental values correspond to HBAR devices operating at room temperatures. In this limit, the performance of HBAR structures is limited by the material attenuation effects~\cite{dutoit1972ultrasonic,dutoit1974microwave}. In our study, we ignore the material attenuation effects and only discuss the diffraction-limited performance. Our study will correspond to devices operating in the cryogenic or potentially milli Kelvin regimes. For example, our results will have high applicability for the HBAR devices that could effectively couple with superconducting qubit systems, etc., which typically operate in the milli Kelvin regime. 

\begin{figure} [htbp]
\includegraphics[width=1.0\linewidth]{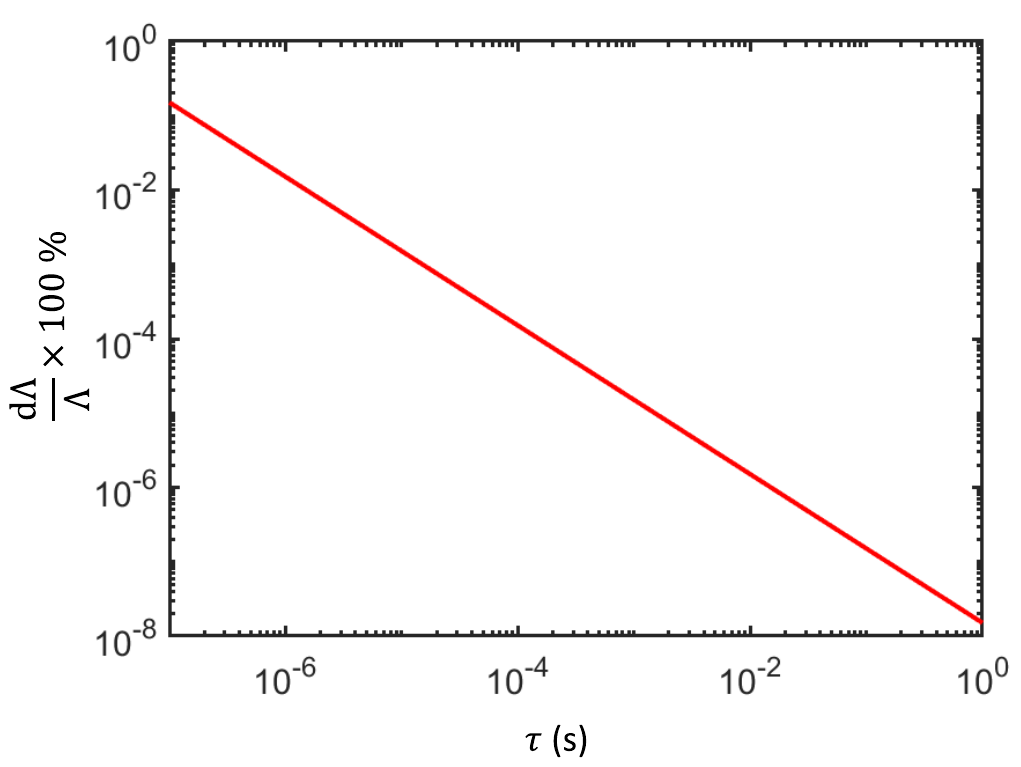}
\caption{{\bf Relative tolerance of $\Lambda$ at various $\tau$} required to predict $\tau$ with $10\%$ accuracy.} 
\label{fig:Sensitivity}
\end{figure}

We discuss here the uncertainties associated with the numerical prediction of the HBAR phonon lifetimes using different methods. In the eigenvalue method, we obtain the lifetimes from the ratio between the HBAR thickness ($t_\text{HBAR}$) and a function of the phonon velocity and the eigenvalue, $\Lambda$, as shown in Eq.~\ref{eq:LifetimeEigVal}. Following Eq.~\ref{eq:LifetimeEigVal}, an error propagation relation can be written as 
\begin{equation}
\left |\frac{\text{d}\tau}{\tau}\right |=\left|\frac{v_2\tau}{t_\text{HBAR}}\times\frac{\text{d}\Lambda}{\Lambda}\right | \propto \left|\tau\times\frac{\text{d}\Lambda}{\Lambda}\right |.
\label{eq:error-tau}
\end{equation}
The uncertainty $\left |\frac{\text{d}\tau}{\tau}\right |$ increases monotonously with $\tau$ when other factors are fixed. It can be deduced from Eq.~\ref{eq:error-tau} that to predict a lifetime within $x\%$ of variability, $\Lambda$ has to be predicted to be within $\frac{t_\text{HBAR}x}{v_2\tau}\%$ of variability. Here, the order of magnitude of the prefactor $\frac{v_2}{t_\text{HBAR}} \sim 10^7$. The prefactor remains fixed since we keep the material and the HBAR thickness unaltered. This implies that we have to be increasingly more stringent with our $\Lambda$ convergence criteria for high lifetime calculations. Figure~\ref{fig:Sensitivity} shows the conditions for desired accuracy (relative tolerance) needed for $\Lambda$ predictions for high-$\tau$ (high-$Q$) systems. When $\tau=0.1\mu$s and the desired accuracy is set to $10\%$, $\Lambda$ must be predicted within $15\%$ accuracy, which is attainable following our numerical procedure. % Considering the challenges we have outlined earlier regarding $\Lambda$ convergence using different methods, it is easy to understand that making high accuracy numerical predictions is highly challenging. 
The accuracy requirement increases fast when the lifetime values are much higher. For example, when $\tau=1$ms, and the desired accuracy is set to $10\%$ of $\tau$ value,  $\Lambda$ must be predicted within ${\sim}10^{-3}\%$ of accuracy. For our $R_{\text{YIG}}=200\mu$m HBAR structure, we continue the simulation till we obtain $\left |\frac{\text{d}\tau}{\tau}\right |=7.9\times10^{-4}$ with $\left |\frac{\text{d}\Lambda}{\Lambda}\right |=4.7\times10^{-7}$, between the last two restarts. On the other hand for the HBAR with $R_{\text{YIG}}=10\mu$m, we continue till $\left |\frac{\text{d}\tau}{\tau}\right |=3.9\times10^{-7}$ with $\left |\frac{\text{d}\Lambda}{\Lambda}\right |=3.8\times10^{-8}$. However, achieving such high accuracy requires high computational expense associated with simulating a large number of iterations. We implement and use an alternative HT approach to expedite the numerical analysis. 

% all planar HBAR cases within a-FBPM. And all for all cases the uncertainty in $\tau$ predictions is less than 1\% between last two iterations, last two restarts and even between mesh densities $N_x=$1024 and 2400. The last one only for 200 and 10 um. Just to give you an idea: for 200um system, $\left |\frac{\text{d}\tau}{\tau}\right |=7.9\times10^{-4}$ with $\left |\frac{\text{d}\Lambda}{\Lambda}\right |=4.7\times10^{-7}$ between last two restarts. Whereas for the 10um system, $\left |\frac{\text{d}\tau}{\tau}\right |=3.9\times10^{-7}$ with $\left |\frac{\text{d}\Lambda}{\Lambda}\right |=3.8\times10^{-8}$. I don't think we need to include all these details.

\subsubsection{Magnon-Phonon Coupling in Diffraction-Limited Regime}

We now turn to discuss the effect of diffraction on the magnon-phonon coupling strength, $g_{mb}$. We estimate the effect of diffraction by computing the ratio between the coupling strength in the planar HBAR structures, $g_{mb}$, and that in the zero-diffraction limit, $g^0_{mb}$. In Fig.~\ref{fig:GmbandTauvsRad}(b)(red), we show the variation of the ratio, $g_{mb}/g^0_{mb}$, with decreasing YIG radius. As expected, The ratio is equal to 1 for low-diffraction cases with $R_{\text{YIG}} \gtrsim 100 \mu$m. As we reduce $R_\text{YIG}$ to $40\mu$m, $g_{mb}$ only changes by 5\% from the $g^0_{mb}$ limit. Note that even though this case typically corresponds to a strong-diffraction regime, the effect of diffraction on $g_{mb}$ is minimal. This is because the YIG disc helps to localize the beam and thus, preserve the magnon-phonon overlap and the coupling strength. 
% This is an improvement in the integration density of $\sim$25-fold compared to the $R_\text{YIG}=200\mu$m structure considered here and a $\sim$100-fold compared to the structure considered in the past research~\cite{xu2021coherent}. 
However, as we decrease $R_\text{YIG} \leq 40 \mu$m, we observe a sharp decline of $g_{mb}$. The percentage of the ratio, $g_{mb}/g^0_{mb}$, drops to $77.4\%$ and $55.7\%$ for HBAR with $R_\text{YIG}=20\mu$m and $10\mu$m, respectively. The decrease of $g_{mb}$ can be explained in the following way. In the zero diffraction limit, the acoustic beam is completely localized in the volume $V_{in}$, of a cylinder with radius equal to $R_{\text{YIG}}$ and the thickness of the HBAR structure. This results in a complete overlap of lateral mode profiles of magnon and phonon modes. Also, the acoustic energy of a propagating beam is completely contained in the volume $V_{in}$, $E_{tot}=E_{in}$. However, as we decrease $R_{\text{YIG}}$, the modeshape spreads out beyond $V_{in}$, due to diffraction. As a result, there is reduced overlap between the phonon and the magnon mode profiles in the YIG region. Correspondingly, some acoustic energy also leaks out of $V_{in}$ and spreads into the GGG region. We refer to the energy of the beam lying outside $V_{in}$ as $E_{out}$. Figure~\ref{fig:GmbandTauvsRad}(b) (black) shows that the ratio $E_{out}/E_{tot}$ increases in the high-diffraction regime, with decreasing $R_\text{YIG}$. 

We combine the magnon-phonon coupling strength (Fig.~\ref{fig:GmbandTauvsRad}(b)) and the phonon lifetimes (Fig.~\ref{fig:GmbandTauvsRad}(a)) to determine the performance figure of merit of the HBAR structures, defined by the magnon-phonon cooperativity, $C=4g^2_{mb}/\kappa_m\kappa_b=4g^2_{mb}\tau_m\tau_b$. We show the variation of $C$ with $R_\text{YIG}$ in Fig.~\ref{fig:GmbandTauvsRad}(c). We assume the magnon lifetime to be $\tau_m=0.07\mu$s~\cite{an2020coherent,xu2021coherent}. We compute the cooperativity values using $\tau_b$ predicted from eigenvalue method (red line), exponential fitting (blue and purple lines), and clipping methods (green line), respectively. We obtain a monotonically decreasing diffraction-limited $C$ ranging from $21.9\times10^4$ to $46.4$, using $\tau$ from the eigenvalue method, as we decrease $R_\text{YIG}$ from $200\mu$m to $10\mu$m. On the other hand, $C$ is in the range between $1.2\times10^4$ and $1.4$, predicted using $\tau$ from the clipping method. As can be noted from Fig.~\ref{fig:GmbandTauvsRad}, $\tau$, $g_{mb}$, and $C$ predicted  by the HT method are in an excellent agreement with a-FBPM predictions, for all $R_\text{YIG}$ cases considered. The close match validates our implementation of the HT method. As we have discussed earlier, a-FBPM has broader applicability compared to HT method, and can be applied for systems with anisotropic material properties and transducer shapes. However, we find that it is an expensive and uncertain process to converge the residual since the rates of convergence for different HBAR structures are slow and variable. Also, one needs to select the number of round trips before restarting the simulation using a trial-and-error approach. In comparison, HT method is computationally inexpensive to make predictions. Additionally, it can be applied to the YIG/GGG HBAR systems since they are isotropic with YIG transducer being a circular thin film disc. Henceforth, we only present predictions from the HT method in this article.  

Note that our calculations predict high cooperativity for even the HBAR with $R_\text{YIG}=10\mu$m. This implies that if material acoustic attenuation effects are eliminated (e.g. by operating in the milli Kelvin regime), one can achieve high cooperativity for planar HBAR structures. However, the phonon lifetime of the HBAR structure with $R_\text{YIG}=10\mu$m is limited to $10.5\mu$s ($0.31\mu$s), as predicted by the eigenvalue (clipping) methods and the maximum magnon-phonon coupling strength is only $55.7\%$ of $g^0_{mb}$. We explore design approaches to further improve these performance parameters. 

\subsubsection{Enhancing integration density and Performance in Diffraction-Limited Regime}

\begin{figure} [htbp]
\includegraphics[width=1.0\linewidth]{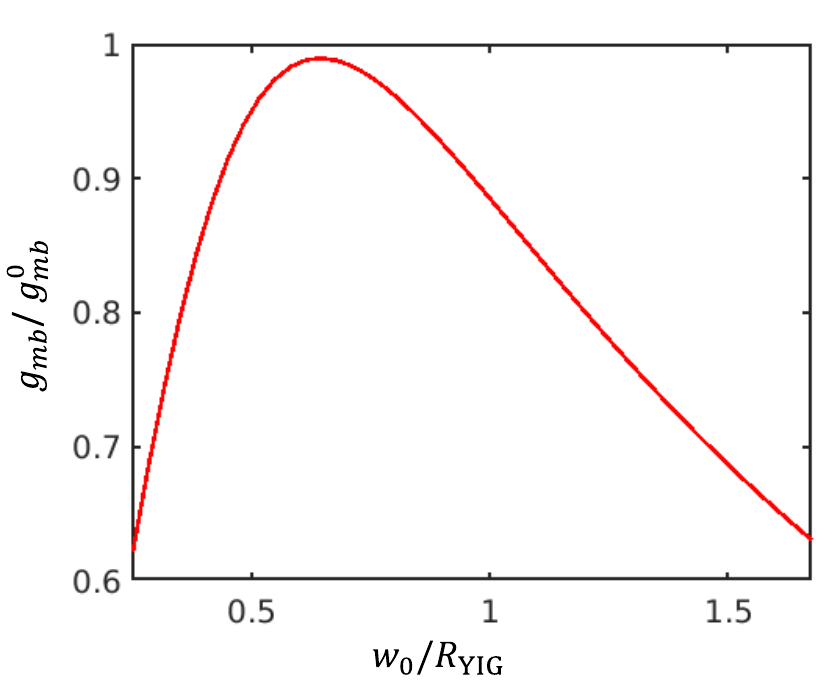}
\caption{{\bf Variation of magnon-phonon coupling, $g_{mb}/g^0_{mb}$, with varying waist of fundamental phonon mode:} $g_{mb}/g^0_{mb}$ is maximum when $w_0/R_\text{YIG}~\sim0.65$. }
\label{fig:TheoreticalLGOverlap}
\end{figure}
% The discussions presented in the previous subsections show that the phonon lifetimes are reduced by three orders or magnitude in the high-diffraction regime, when $R_{\text{YIG}} = 10 \mu$m, compared to low-diffraction cases. The magnon-phonon coupling is also reduced by a factor of two. The phonon lifetime reduction plays a dominant role and results in a three orders of magnitude reduction of the cooperativity, the performance figure of merit of the HBAR structures. Here, we discuss a strategy to mitigate the diffraction losses and improve the performance of the HBAR structures. 
Here, we illustrate that the use of focusing dome-like surface structures could significantly improve the performance of HBAR structures. Past studies demonstrated that the confocal HBAR structures could achieve Q-factors on the order of $10^7$, while resulting in $10^3$-fold reduction in device volumes~\cite{kharel2018ultra}. However, such structures have not been explored for hybrid magnomechanical systems. We show that both phonon lifetimes and magnon-phonon coupling can be improved by employing confocal geometries, leading to improved performance of hybrid magnomechanical systems. 
% due to lowered mode volume of phonons resulting from reduced lateral spread of the phonon modes. 
\begin{figure*} [htbp]
\includegraphics[width=1.0\linewidth]{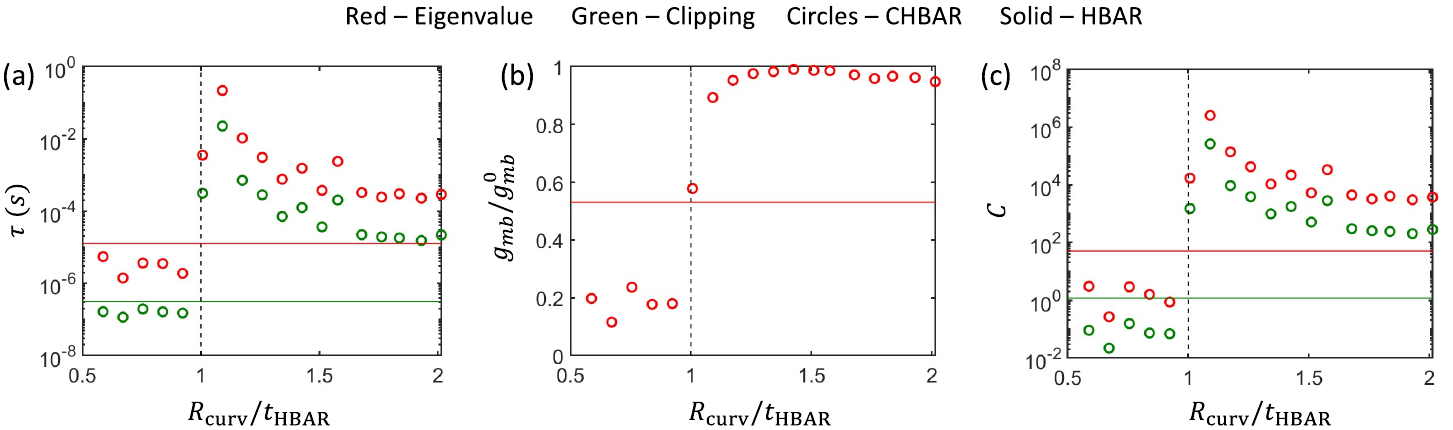}
\caption{{\bf Performance of CHBAR structures with varied radius of curvature, $R_\text{curv}$, of the dome-shape:} (a) Phonon lifetimes, $\tau$, calculated from the eigenvalue (red) and clipping methods (green), following the HT method, (b) Ratio between magnon-phonon coupling in CHBAR structures and that in the zero-diffraction limit, $g_{mb}/g^0_{mb}$ and (c) Magnon-phonon cooperativity, $C$. $R_\text{curv}$ is varied keeping $R_\text{YIG}=10\mu$m and $R_\text{cross}=60\mu$m fixed. Solid lines represent corresponding values for a planar HBAR structure with $R_\text{YIG}=10\mu$m, while the circles of the same color correspond to the CHBAR values.}
\label{fig:TauGmbvsRCurv}
\end{figure*}
We first identify the optimal domeshape that will result in the maximum overlap between the CHBAR phonon and the magnon modeshapes, and consequently, the highest $g_{mb}$. We obtain theoretical estimates of the overlap and identify the dome shape where the overlap is maximum. We assume that the phonon modeshapes in CHBAR structures can be described by Laguerre-Gaussian (LG) functions. The fundamental mode, LG$_{00}$, can be assumed to be of the form: $u_y \propto e^{-R^2/w_0^2}$, due to the axi-symmetry of our system. Here, $R$ is the radial coordinate $\sqrt{x^2+y^2}$. $w_0$ is the waist of the Gaussian beam at $z=0$ (see Fig.~\ref{fig:HBARConfigs}), that can be varied by changing $t_\text{HBAR}$, $\omega$, or the dome radius of curvature, $R_\text{curv}$. We normalize the modeshapes according to Eqs.~\ref{eq:MagnonNormal},~\ref{eq:PhononNormal} and calculate the coupling strengths using Eq.~\ref{eq:MagnonPhonon}. 
% $J_0(\frac{R}{R_\text{YIG}}\zeta_0)$? is the magnon modeshape, which is also the phonon excitation function. But the phonon modeshape itself just depends on the cavity not on the excitation function. Here, phonon modeshape is a Gaussian while magnon modeshape is a Bessel function.
Figure~\ref{fig:TheoreticalLGOverlap} shows the computed $g_{mb}/g^0_{mb}$ ratio as a function of the beam waist, $w_0$. We find that the peak of $g_{mb}/g^0_{mb}$ occurs at $w_0/R_\text{YIG}~\sim0.65$. $g_{mb}$ decreases sharply if $w_0 \lesssim 0.65 R_\text{YIG}$, however, decreases slowly if $w_0$ is increased beyond the optimal value. Interestingly, we find that $w_0/R_\text{YIG}$ is a constant and does not depend on $R_\text{YIG}$, for all the CHBAR structures considered in this article. We use the $w_0$ value to obtain an initial estimate of the radius of curvature, $R_\text{curv}$, of the dome~\cite{newberry1989paraxial}: 
\begin{subequations}
\label{eq:RcurvEstimate}
\begin{align}
        R_\text{curv} &= \frac{1}{Re[1/(q_0+t_\text{HBAR})]},\\
         \text{with  } q_0 &= \frac{i\pi w_0^2}{\lambda}.
\end{align}
\end{subequations}
We obtain the analytical estimate to be $R_\text{curv}/t_\text{HBAR}{\sim}1.5$ that corresponds to $w_0/R_\text{YIG}~{\sim}0.65$ and the peak of $g_{mb}/g_{mb}^{0}$ as shown in Fig.~\ref{fig:TheoreticalLGOverlap}. Note that $R_\text{curv}$ can be similarly estimated for various device thicknesses ($t_\text{HBAR}$), wavelengths ($\lambda$) and waists by appropriately accounting for these variables shown in Eq.~\ref{eq:RcurvEstimate}. This $R_\text{curv}$ estimate results in maximum $g_{mb}/g^0_{mb}$, however, this analysis does not consider the effects of the dome shape on phonon lifetimes and cooperativities. Also, we ignore the effects of the lateral extent of the HBAR structure, the localizing effects of the YIG film, and assume phonon modeshapes to be perfectly Gaussian. 

\begin{figure*} [htbp]
\includegraphics[width=1.0\linewidth]{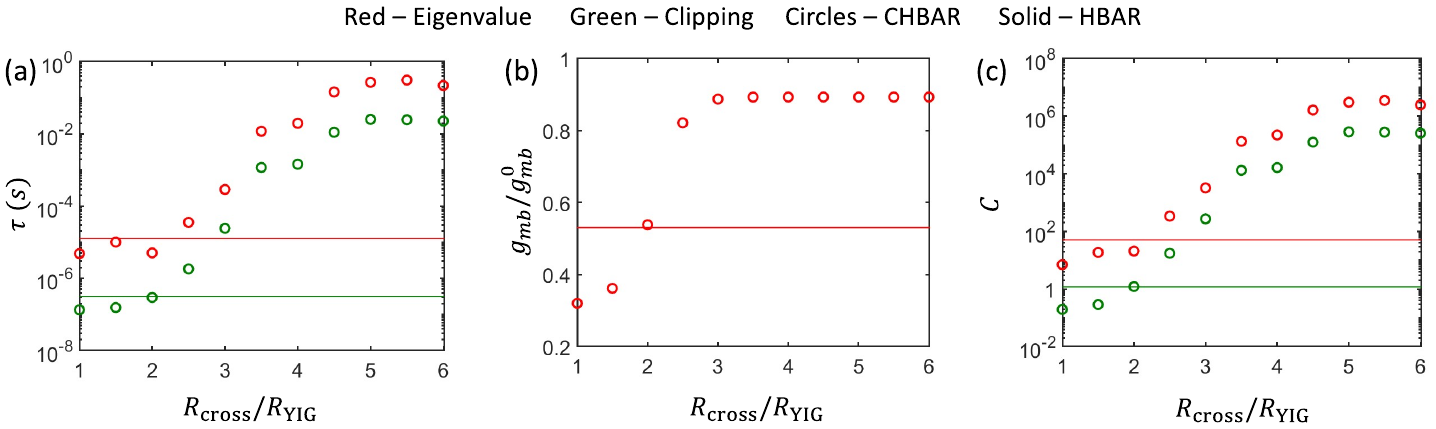}
\caption{{\bf Performance of CHBAR structures with varied radius of cross-section, $R_\text{cross}$, of the dome-shape:} (a) Phonon lifetimes, $\tau$, calculated from the eigenvalue (red) and clipping methods (green), following the HT method, (b) Ratio between magnon-phonon coupling in HBAR structures and that in the zero-diffraction limit, $g_{mb}/g^0_{mb}$ and (c) Magnon-phonon cooperativity, $C$. $R_\text{cross}$ is varied keeping $R_\text{YIG}=10\mu$m and $R_\text{curv}/t_{\text{HBAR}}=1.09$ fixed. Solid lines represent corresponding values for a planar HBAR structure with $R_\text{YIG}=10\mu$m, while the circles of the same color correspond to the CHBAR values.}
\label{fig:TauGmbvsRCross}
\end{figure*}

To obtain a $R_\text{curv}$ estimate that improves the overall performance of these CHBAR structures, we carry out a systematic numerical analysis using the HT method. We modify the reflection operator at the upper GGG surface to include the effects of the lateral extents of the dome and the device, the attenuation window, and the localization effects of the YIG film. We include the phase induced by the dome surface in Eq.~\ref{eq:ReflectionGGG} and the radial form of Eq.~\ref{eq:HKHBARreflection} and write the modified reflection opertor as 
\begin{equation}
    R_{t_\text{CHBAR},m}(x,y) = 
    \begin{cases}
        e^{ik_{z0,m}r^2/R_\text{curv}},& \text{if } r\leq R_\text{cross}\\
        1,& \text{if } R_\text{cross}<r\leq \frac{W_\text{eff}}{2}\\
        0.              & \text{otherwise}
    \end{cases}
    \label{eq:ReflectionDome}
\end{equation}
We vary $R_\text{curv}/t_\text{HBAR}$ across the analytical estimate, ${\sim}1.5$, and identify $R_\text{curv}$ that results in the optimal performance of the CHBAR structures. We show the effect of the focusing dome on the phonon lifetime, magnon-phonon coupling and cooperativity of the CHBAR structure in Fig.~\ref{fig:TauGmbvsRCurv}(a), (b) and (c), respectively. We show the predictions from the eigenvalue method and the clipping method using red and green circles, respectively. We also show the corresponding performance parameters of the planar HBAR structure with red and green horizontal lines, respectively. For the clipping method, we consider the volume, $V_{in}$, defined by the dome's lateral cross-sectional area and the CHBAR thickness. We fix $R_\text{YIG}=10\mu$m and the dome $x-y$ cross-section radius, $R_\text{cross}=60\mu$m, for this analysis. As can be seen from Fig.~\ref{fig:TauGmbvsRCurv}(a), the phonon lifetimes are maximum at $R_\text{curv}/t_\text{HBAR}=1.09$. This value is less than the analytical prediction for maximum $g_{mb}$, $R_\text{curv}/t_\text{HBAR} \sim 1.5$. However, the peak value, $\tau{\sim} 218$ms (red circles), is ${\sim}$500 times larger than the value at $R_\text{curv}/t_\text{HBAR}{\sim}1.5$, justifying the necessity of performing the systematic analysis. The $\tau$ peak value for the CHBAR structure is ${\sim}1.7\times10^4$ (red circles) (${\sim}7.3\times10^4$, green circles) times greater than the corresponding planar HBAR $\tau$ prediction, shown with red (green) horizontal lines, respectively. The peak $\tau$ predicted by the clipping method is ${\sim} 22.7$ms, lower than those predicted by the eigenvalue method, as expected. We note that $\tau$ decrease sharply for $R_\text{curv}/t_\text{HBAR} \text{\textless} 1$. This corresponds to the situation when the center of curvature of the dome is inside the HBAR structure, resulting in a negative `$g-$parameter'~\cite{kharel2018ultra}, where $g=1-\frac{t_\text{HBAR}}{R_\text{curv}}$. It is shown that one cannot obtain real and finite Gaussian beam solutions for structures with a negative $g-$parameter~\cite{kharel2018ultra}. We find that this is the case for the modes corresponding to $R_\text{curv}/t_\text{HBAR}\text{\textless}1$ in our case, that is, they are lossy modes which do not have well-defined Gaussian characters. 

Figure~\ref{fig:TauGmbvsRCurv}(b) shows that the coupling, $g_{mb}/g^0_{mb}$, is approximately 0.99 at $R_\text{curv}/t_\text{HBAR}{\sim}1.4$, a value close to the analytical estimate of ${\sim} 1.5$. These results show that reducing the phonon mode volume by means of focusing does not necessarily improve $g_{mb}$ beyond the maximum achievable value, $g^0_{mb}$, which represents the case when there is complete lateral overlap of phonon and magnon modes. 
% However, $\tau$ is reduced by more than two orders of magnitude at this point. 
We use $\tau$ and $g_{mb}$ to calculate the figure of merit, $C$, and show the results in Fig.~\ref{fig:TauGmbvsRCurv}(c). The peak $C$ value is predicted to be $2.5\times10^6$ ($2.6\times10^5$), using the eigenvalue (clipping) method. The $C$ values reach peak at $R_\text{curv}/t_\text{HBAR}=1.09$, where $\tau$ is maximum, as we have noted from Fig.~\ref{fig:TauGmbvsRCurv}(a). The $\tau$ and $C$ values of the CHBAR structure are several orders of magnitude improved compared to its planar counterpart and the coupling reaches up to $90\%$ of $g^0_{mb}$. This result highlights that $\tau$ is the dominating factor that controls the performance of CHBAR structures, since $g_{mb}$ is always limited to the maximum value, $g_{mb}^0$. Note that we only vary $R_\text{curv}$ for this analysis and keep the radius of the dome cross-section, $R_\text{cross}$, fixed at $60\mu$m. Due to the fixed cross-sectional area of the dome, the lateral integration density of these devices, $D_i=\frac{A^0_d}{A_d}$, is limited to $\sim 64$. Here, $A_d$ is the cross-sectional area of the confocal dome surface and $A^0_d = 0.72$ mm$^2$~\cite{xu2021coherent}.

To further improve the integration density, we analyze CHBAR structures with reduced dome cross-sectional area. We keep $R_\text{curv}/t_\text{HBAR}=1.09$ and $R_\text{YIG}=10\mu$m fixed for this analysis. Figure~\ref{fig:TauGmbvsRCross} (a), (b) and (c) show the variation of $\tau$, $g_{mb}$, and $C$ with $R_\text{cross}/R_\text{YIG}$, respectively. As we decrease $R_\text{cross}/R_\text{YIG}$ from $6$ to $1$, $\tau$ is reduced by more than 4 orders of magnitude from ${\sim}218$ ms (22.7 ms) to $4.83\mu$s ($0.134\mu$s), as predicted by the eigenvalue (clipping) method. $g_{mb}$ shows a sharp decline for $R_\text{cross}/R_\text{YIG} \text{\textless} 3$ reaching the lowest value of $g_{mb}/g^0_{mb}~{\sim}0.32$ at $R_\text{cross}/R_\text{YIG}=1$, however, remains mostly unchanged for $R_\text{cross}/R_\text{YIG}\geq3$. The cooperativity follows a similar trend to that of $\tau$, as shown in Fig.~\ref{fig:TauGmbvsRCross}(c). Using the predictions from the eigenvalue (clipping) method, we obtain that $C$ is decreased by more than 5 orders of magnitude from  $2.5\times10^6$ ($2.6\times10^5$) to 7.2 (0.19) as we decrease $R_\text{cross}/R_\text{YIG}$. Additionally, $\tau$, $g_{mb}$, and $C$ of these CHBAR structures is reduced below those of the planar HBAR values, shown with solid lines, if $R_\text{cross}/R_\text{YIG} \text{\textless}2$. Considering all the different aspects, we argue that the optimal geometry for the CHBAR structures is represented by the condition $R_\text{cross}/R_\text{YIG}=4.5$. Our analysis using the HT eigenvalue (clipping) method predicts that the diffraction-limited lifetime of $\tau=$144.7 ms (11.1 ms) and a integration density of $D_i=113$ could be achieved at a cooperativity $C=1.64\times10^6$ ($1.25\times10^5$) for the CHBAR structure with $R_\text{curv}/t_\text{HBAR}=1.09$, $R_\text{YIG}=10\mu$m and $R_\text{cross}/R_\text{YIG}=4.5$. The results presented in Figs.~\ref{fig:TauGmbvsRCurv} and ~\ref{fig:TauGmbvsRCross} provide a proof-of-concept that including a focusing dome improves the performance of the hybrid magnonic HBAR structures for quantum memory and transduction applications.

% \begin{figure*} [htbp]
% \includegraphics[width=1.0\linewidth]{Figures/LifetimeandGmbEffect.pdf}
% \caption{Effect of (a). improved $\tau$ and (b) improved $g_{mb}$ on the pulse echo amplitudes $a$ of the microwave photon.}
% \label{fig:LifetimeandGmbEffect}
% \end{figure*}

% \begin{figure} [htbp]
% \includegraphics[width=1.0\linewidth]{Figures/200um_TiltEffect.png}
% \caption{(a). (b). (c).}
% \label{fig:200umTiltEffect}
% \end{figure}

\section{Conclusion and Outlook}

In this article, we establish a modeling approach that can be broadly used to design hybrid magnonic HBAR structures for high-density, long-lasting quantum memories and efficient quantum transduction devices. We illustrate the approach by discussing the magnon-phonon transduction properties of hybrid magnonic YIG/GGG HBAR structures. We present analytical and numerical analyses of the bulk acoustic wave phonon mode lifetimes, and the magnon-phonon coupling strengths in planar and confocal YIG/GGG HBAR structures. We discuss strategies to improve the phonon mode lifetimes of these structures, since increased lifetimes have direct implications on the storage times of quantum states for quantum memory applications. Additionally, high integration density of on-chip memory or transduction centers is naturally desired for high-density memory or transduction devices.  
% Additionally, high integration density is naturally desired to accommodate multiple on-chip memory or transduction centers or other on-chip hardware alongside the HBAR structures. 
In our structures, the transduction centers are represented by the YIG films and thus, integration density can be increased by reducing the lateral dimension of these films. However, the reduced aperture results in high-diffraction that affects the performance of these devices. We analyze the diffraction-affected shear wave phonon modes in the HBAR structures using two different methods: (1) Fourier beam propagation (FBPM) and (2) Hankel transform (HT) eigenvalue problem method. The FBPM method has been widely used to analyze beam propagation in the field of optics, and more recently, to study HBAR phonons in planar and confocal HBAR (CHBAR) structures. 
% The advantage of FBPM lies in its simplicity, broad applicability and the ability to predict the field profiles at any target distances from the source. Here, we implement a reformulated approach that allows us to achieve a seven-fold speed up of the computation time. 
% However, we find that the it is an expensive and uncertain process to converge the modes using the FBPM approach. 
We find that the FBPM method often requires us to analyze large and unpredictable number of round trips to obtain converged phonon modes. We implement an adaptive FBPM (a-FBPM) approach that mostly overcomes the slow convergence issues. However, we find that a-FBPM still suffers from convergence challenges when applied to confocal HBAR structures. To circumvent the challenges, we implement the HT method that allows us to obtain converged phonon modes with significantly reduced computational cost. The HT method leverages the axi-symmetry of the problem and the isotropic material properties of the YIG/GGG HBAR structures to reduce the 3D problem to a 2D one and expedite the analysis. The HT method has been mostly used in the field of optics, e.g., Fabry-Perot cavities, however, it has not been applied for acoustics analysis, to the best of our knowledge. 

Our study provides key insights into the diffraction-limited performance of the YIG/GGG HBAR structures. We predict the diffraction-limited $\tau$  to be on the order of milliseconds, for a planar HBAR structure with lateral YIG dimension, $R_\text{YIG}=200\mu$m. A recent study reported that the performance of a YIG/GGG HBAR structure at room temperature is limited by the phonon lifetime at $0.25\mu$s~\cite{xu2021coherent}. The previously studied structure had larger YIG lateral area than our structures and thus, the diffraction effects were less dominant. The phonon lifetime in HBAR structures could be limited by both material and diffraction losses depending on the device geometry and/or the operating conditions. At room temperature, the phonon lifetime is primarily limited by acoustic attenuation effects. However, these effects are less significant at low temperatures while the diffraction losses play an important role. Therefore, our results will have high applicability for devices operating in the cryogenic or potentially milliKelvin (mK) regimes. For example, our approach and analyses can be applied to design HBAR devices that could effectively couple with superconducting qubit systems. We acknowledge that the analysis of the material losses is necessary to obtain a complete understanding of the performance of the YIG/GGG systems. This may include an understanding of the different attenuation effects (e.g., Akhiezer and Landau-Rumer) at various temperatures and frequencies of interest, and strategies to overcome them. 

Assuming that the material-limited lifetimes could be increased to $~{\sim} 0.1$ ms at mK temperatures, we find that the planar HBAR structures are not affected by diffraction effects even at $R_\text{YIG}=50\mu$m,. This structure already offers a significant 50-fold improved integration density over the reference structure~\cite{xu2021coherent}. The integration density can be further improved by scaling down the YIG film lateral area. However, the reduced aperture will affect the phonon lifetime and the magnon-phonon coupling strength, resulting in a decrease of the cooperativity, the performance figure of merit for magnon-phonon transduction. It is therefore imperative to develop a strategy that increases the integration density of HBAR structures without affecting the performance. Additionally, the performance of planar HBAR structures was found to be highly sensitive to the parallel nature of the surfaces. To address both these aspects, we investigate confocal YIG/GGG HBAR structures with top focusing domes and a planar bottom surfaces. Use of focusing dome structures has been proposed to significantly improve the phonon lifetime and integration density of these systems. Furthermore, the dome structures eliminate the necessity of maintaining perfectly parallel surfaces of planar HBAR structures. We first theoretically estimate the shape of the dome structure for which the magnon-phonon coupling strength, $g_{mb}$, is at its maximum value. 
% The coupling is highest when there is maximum overlap between the CHBAR phonon modeshapes with magnon modes. We theoretically estimate the overlap and extract the shape paramters using the condition where the overlap is maximum.
% We assume that the fundamental phonon mode can be described by a Laguerre-Gaussian (LG) function. The waist of the Gaussian beam at the bottom surface of the HBAR structure is an adjustable parameter that can be varied by changing the dome radius of curvature, HBAR thickness, or phonon mode frequency. 
% % We ignore the effects of lateral extent of the structure and the dome, and the localizing effects of the YIG film. 
% We find that the magnon-phonon coupling peaks at a Gaussian beam waist-to-YIG film radius ratio of 0.65. We obtain analytical estimate for the radius of curvature from the optimal waist and the HBAR thickness. The estimated radius of curvature results in maximum the magnon-phonon coupling, however, does not discuss the effect of the dome shape on phonon lifetime or cooperativity. Also, the analytical estimate ignores the effects of YIG film localization. We obtain a more accurate estimate using the HT approach. 
We then perform a rigorous numerical analysis and obtain a refined set of shape parameters of the dome for which both $\tau$ and $C$ are optimal, and $g_{mb}$ is close to its peak value. 
% We perform additional numerical analysis to identify optimal radius of cross-section of the dome as well. 
Overall, we find that ultra-high, diffraction-limited, cooperativities and phonon lifetimes on the order of ${\sim}10^5$ and ${\sim}10$ ms, respectively, could be achieved using a CHBAR structure with $R_\text{YIG}=10\mu$m. In addition to enhanced $\tau$ and $C$, the confocal HBAR structure will offer more than 100-fold improvement of integration density. 

% However, the study shows that magnon-phonon coupling strengths $g_{mb}$ cannot be improved by focusing the phonon and reducing its mode volume using a CHBAR structure, which however helps in recovering ${\sim}90\%$ of $g^0_{mb}$ in the high-diffraction regime. This is in contrast with the general understanding that reducing the phonon mode volumes can help increase its coupling strengths with other quantum information carriers, such as photons (optomechanics) and spin qubits. Following this analysis, the dynamics of the triply resonant system is analyzed employing the Heisenberg-Langevin formalism and insights are provided towards the effect of enhanced $\tau$ and $g_{mb}$ on the coherent pulse echo behavior. It is found that improving $\tau$ results in sustained pulse echoes. Whereas increasing $g_{mb}$, while results in strong pulse echoes initially, also increases the decay of these echo peaks and also increases spurious oscillations in between the peaks.  

In this work, we discuss the diffraction-limited performance induced by the lateral area of the YIG film and the dome structure. It will be interesting to apply the insights presented in the article and explore the applicability of YIG/GGG HBAR structures for quantum transduction applications, as future work. For example, these HBAR structures could be used for coupling with other quantum information carriers such as superconducting qubits, which operate in the microwave frequency regime and at mK temperatures. However, to optimize the performance of such devices, a further comprehensive analysis of different geometric and physical parameters will be necessary. For example, we keep the thickness fixed at $527.2\mu$m for all YIG/GGG HBAR structures investigated in this work. The same thickness allows us to keep the free spectral range of phonons fixed for all our analysis. The different geometric parameters, such as the overall thickness, the thickness of the YIG and the GGG regions, geometrical misalignment, and imperfections, or, the physical parameters, such as the photon and magnon lifetimes and their coupling strengths, could also play important role in the overall device performance. We assume that these parameters remain invariant in our analysis. A comprehensive optimization analysis can be performed using the current state-of-the-art machine learning techniques, which is a promising research direction for the future. Additionally, we only discuss the coupling between the fundamental magnon mode and the fundamental phonon modes of the planar and the confocal HBAR structures. It will be interesting to consider the effect of higher-order mode couplings in the device performance. The higher-order mode couplings can be particularly relevant here since these structures host long-lasting phonon modes with broadband nature. Other interesting directions to explore are the anharmonic phonon interactions at the surface and interfaces, and the coupling of bulk acoustic waves and the surface acoustic waves. Our study is concerned with coherent states, and it will be important to explore how the insights provided here translate to systems involving non-classical states, e.g. squeezed states, cat states, and Fock states. 

\section{Acknowledgements}

This work was partially supported by funding from the Quantum Explorations in Science \& Technology (QuEST) grant provided by the University of Colorado Boulder Research \& Innovation Office in partnership with the College of Engineering and Applied Science, the College of Arts and Sciences, JILA, and the National Institute of Standards and Technology (NIST). We acknowledge the computing resources provided the RMACC Summit supercomputer, which is supported by the National Science Foundation (awards ACI-1532235 and ACI-1532236), the University of Colorado Boulder, and Colorado State University. The Summit supercomputer is a joint effort of the University of Colorado Boulder and Colorado State University.

\bibliography{mpLiterature}

\end{document}